\pdfoutput=1
\documentclass[11pt,twoside,a4paper,cmspaper,final,collab]{cms-tdr}
\def\svnVersion{89130b0}\def\svnDate{2024/05/17}\def\cmsCernNoTag{CERN-EP-2024-025}\def\cmsCernDate{\today}\def\cmsMessage{Published in the Journal of High Energy Physics as \href{http://dx.doi.org/10.1007/JHEP05(2024)047}{\doi{10.1007/JHEP05(2024)047}.}}
\begin{document}\cmsNoteHeader{EXO-23-014}

\newcommand{\tkisoSTA}{\ensuremath{I^\text{STA}_\text{rel, trk}}\xspace}
\newcommand{\tkisoTMS}{\ensuremath{I^\text{TMS}_\text{rel, trk}}\xspace}
\newcommand{\DeltaR}{\ensuremath{\Delta R}\xspace}
\newcommand{\mMuMu}{\ensuremath{m_{\PGm\PGm}}\xspace}
\newcommand{\mMuMuCorr}{\ensuremath{m^\text{corr}_{\PGm\PGm}}\xspace}
\newcommand{\vecptMuMu}{\ensuremath{\vec{p}_{\text{T},\PGm\PGm}}\xspace}
\newcommand{\Lxy}{\ensuremath{L_{\text{xy}}}\xspace}
\newcommand{\vecLxy}{\ensuremath{\vec{L}_{\text{xy}}}\xspace}
\newcommand{\LxyErr}{\ensuremath{\sigma_{\Lxy}}\xspace}
\newcommand{\LxySig}{\ensuremath{\Lxy/\LxyErr}\xspace}
\newcommand{\dzero}{\ensuremath{d_{\text{0}}}\xspace}
\newcommand{\dzeroErr}{\ensuremath{\sigma_{\dzero}}\xspace}
\newcommand{\dzeroSig}{\ensuremath{\dzero/\dzeroErr}\xspace}
\newcommand{\DeltaPhi}{\ensuremath{\Delta\Phi}\xspace}
\newcommand{\DeltaPhiAbs}{\ensuremath{\abs{\Delta\Phi}}\xspace}
\newcommand{\DeltaPhiAbsC}{\ensuremath{\abs{\Delta\Phi}_\text{C}}\xspace}
\newcommand{\DetaMuMu}{\ensuremath{\abs{\Delta \eta _{\PGm\PGm}}}\xspace}

\newcommand{\PSMHiggs}{\ensuremath{\PH}\xspace}
\newcommand{\PDarkHiggs}{\ensuremath{\PH_\text{D}}\xspace}
\newcommand{\PZD}{\ensuremath{\PZ_\text{D}}}
\newcommand{\PLLP}{\ensuremath{\text{LLP}}}
\DeclareRobustCommand{\PAf}{{\HepAntiParticle{\Pf}{}{}}\Xspace}
\newcommand{\PRPVSquark}{{\ensuremath{\PSQ}}\xspace}
\newcommand{\PRPVChi}{\ensuremath{\PSGczDo}\xspace}
\newcommand{\squarktoqchi}{\mbox{$\PRPVSquark \to \PQq\PRPVChi$}\xspace}
\newcommand{\MuMu}{\ensuremath{\PGmp\PGmm}\xspace}
\newcommand{\MuMuNu}{\ensuremath{\MuMu\PGn}\xspace}
\newcommand{\chitoeenu}{\ensuremath{\PRPVChi\to\Pep\Pem\PGn}\xspace}
\newcommand{\chitomumunu}{\ensuremath{\PRPVChi\to\MuMuNu}\xspace}
\newcommand{\mSquark}{\ensuremath{m(\PRPVSquark)}\xspace}
\newcommand{\mChi}{\ensuremath{m(\PRPVChi)}\xspace}

\newcommand{\mh}{\ensuremath{m(\PSMHiggs)}\xspace}
\newcommand{\mZD}{\ensuremath{m(\PZD)}\xspace}
\newcommand{\cTau}{\ensuremath{c\tau}\xspace}

\renewcommand{\Lone}{L1\xspace}

\newlength\DFigWidth
\setlength\DFigWidth{0.77\textwidth}
\newlength\DSquareWidth
\setlength\DSquareWidth{0.51\textwidth}

\newlength\cmsTabSkip\setlength{\cmsTabSkip}{1ex}

\cmsNoteHeader{EXO-23-014}

\title{Search for long-lived particles decaying to final states with a pair of muons in proton-proton collisions at \texorpdfstring{$\sqrt{s} = 13.6\TeV$}{sqrt(s) = 13.6 TeV}}

\date{\today}

\abstract{An inclusive search for long-lived exotic particles (LLPs) decaying to final states with a pair of muons is presented.  The search uses data corresponding to an integrated luminosity of 36.6\fbinv collected by the CMS experiment from the proton-proton collisions at $\sqrt{s} = 13.6\TeV$ in 2022, the first year of Run 3 of the CERN LHC.  The experimental signature is a pair of oppositely charged muons originating from a secondary vertex spatially separated from the proton-proton interaction point by distances ranging from several hundred $\mum$ to several meters. The sensitivity of the search benefits from new triggers for displaced dimuons developed for Run 3. The results are interpreted in the framework of the hidden Abelian Higgs model, in which the Higgs boson decays to a pair of long-lived dark photons, and of an $R$-parity violating supersymmetry model, in which long-lived neutralinos decay to a pair of muons and a neutrino. The limits set on these models are the most stringent to date in wide regions of lifetimes for LLPs with masses larger than 10\GeV.}

\hypersetup{
pdfauthor={CMS Collaboration},
pdftitle={Search for long-lived particles decaying to final states with a pair of muons in proton-proton collisions at sqrt(s) = 13.6 TeV},
pdfsubject={CMS},
pdfkeywords={CMS, Exotica, long-lived particles, dimuons}}

\maketitle 

\section{Introduction} \label{sec:introduction}
Long-lived particles (LLPs) that are unobserved as yet are predicted by many extensions of the
standard model (SM), in particular by various supersymmetry (SUSY)
scenarios~\cite{Hewett:2004nw, Barbier:2004ez} and ``hidden-sector''
models~\cite{Strassler:2006im, Han:2007ae}. 
Such particles could manifest themselves through decays to SM
particles at macroscopic distances from the proton-proton ($\Pp\Pp$)
interaction point (IP).

This paper describes an inclusive search for an exotic massive LLP
decaying to a pair of oppositely charged muons, referred to as a ``displaced dimuon'', originating from a secondary vertex spatially separated from
the IP.  The analysis is based on a data set of $\Pp\Pp$ collisions
corresponding to an integrated luminosity of 36.6\fbinv
collected with the CMS detector at $\sqrt{s} = 13.6\TeV$ during 2022, the first year of Run~3 of the CERN LHC.
It is a continuation and extension of
the CMS analysis~\cite{EXO-21-006} performed using data collected at
$\sqrt{s} = 13\TeV$ during Run~2 (2016--2018) and corresponding to
an integrated luminosity of 97.6\fbinv (referred to below as the CMS Run~2 analysis).
A minimal set of requirements and loose event selection criteria 
allow the search to be sensitive to a wide range of
models predicting LLPs that decay to final states that include a pair of
oppositely charged muons.  
Improvements in the triggers result in a significantly higher efficiency for displaced dimuons in Run~3,
particularly at dimuon masses of a few tens of GeV.
The present search explores the LLP mass range above 10\GeV and
is sensitive to vertex
displacements ranging from several hundred $\mum$ to several meters.

We interpret the results of the search in the
frameworks of two benchmark models: the hidden Abelian Higgs model (HAHM), in
which displaced dimuons arise from decays of hypothetical dark
photons~\cite{Curtin:2014cca}, and a simplified SUSY model, in which long-lived neutralinos decay to a pair of muons and a neutrino as a result of $R$-parity violation (RPV)~\cite{Barbier:2004ez}.
The results for the HAHM model from Run~3 are statistically combined with the
results of the Run~2 analysis~\cite{EXO-21-006}.
The Run~2 search~\cite{EXO-21-006} and the present search
are complementary to a CMS search~\cite{CMS:2021sch}, which uses data collected in Run~2 with a dedicated high-rate data stream
in order to explore otherwise inaccessible parameter space at low dimuon masses.
A search for LLPs decaying to displaced dimuons has also been performed by
CMS in Run~1, using data taken at $\sqrt{s} = 8\TeV$ and corresponding
to an integrated luminosity of 20.5\fbinv~\cite{EXO-12-037,
  CMS-PAS-EXO-14-012}, and by
the ATLAS Collaboration in Run~2, using data corresponding to an
integrated luminosity of 32.9\fbinv~\cite{Aaboud:2018jbr, Aad:2683781}.

This paper is organized as follows.  Section~\ref{sec:detector}
describes the CMS detector.  Section~\ref{sec:samples} presents the
signal models, as well as the simulated signal and background event samples.
 Section~\ref{sec:selection} describes the analysis
strategy, the triggers, and the offline event selection.  Estimation of
backgrounds and the associated systematic uncertainties are described
in Section~\ref{sec:bkg_evaluation}.  Section~\ref{sec:systunc_signal}
summarizes the systematic uncertainties affecting signal
efficiencies.  Section~\ref{sec:results} describes the results
of this analysis
and their combination with the results of the CMS Run~2 analysis.
The summary of the paper is given in Section~\ref{sec:summary}.
Tabulated results and supplementary material for reinterpreting the
results in the framework of models not explicitly considered in this
paper are provided in the HEPData record for this analysis~\cite{HEPData}.

\section{The CMS detector} \label{sec:detector}
The central feature of the CMS detector is a superconducting solenoid
of 6\unit{m} internal diameter, providing a magnetic field of
3.8\unit{T}.  Within the solenoid volume are a silicon pixel and strip
tracker extending outwards to a radius of 1.1\unit{m},
a lead tungstate crystal electromagnetic calorimeter,
and a brass and scintillator hadron calorimeter, each composed
of a barrel and two endcap sections.  Forward calorimeters extend the
coverage in pseudorapidity $\eta$ provided by the barrel and endcap
detectors.  Muons are detected in gas-ionization chambers covering the
range $\abs{\eta} < 2.4$ and embedded in the steel flux-return yoke
outside the solenoid.  The muon system is composed of four types of
chambers: drift tubes (DTs) in the barrel ($\abs{\eta} < 1.2$), cathode strip chambers
(CSCs) in the endcaps ($0.9 < \abs{\eta} < 2.4$), resistive-plate chambers in both the
barrel and the endcaps, and gas electron multipliers in the forward regions of the endcaps.  
The chambers are assembled into four
``stations'' at increasing distance from the IP.
The stations in the barrel are located approximately 4, 5, 6, and
7\unit{m} away from the IP radially, while the stations in the endcap
are located approximately 7.0, 8.0, 9.5, and 10.5\unit{m} away from
the IP along the beam line axis on both ends of the detector.
Each station provides reconstructed hits in several
detection planes, which are combined into track segments, forming the
basis of muon reconstruction in the muon system~\cite{Sirunyan:2018}.
A more detailed description of the CMS detector, together with a
definition of the coordinate system used and the relevant kinematical
variables, can be found in Refs.~\cite{Chatrchyan:2008zzk, CMS:2023gfb}.

Events of interest are selected using a two-tiered trigger system.
The first level (\Lone), composed of custom hardware processors, uses
information from the calorimeters and muon detectors to select events
at a rate of approximately 100\unit{kHz} within a fixed latency of 4\mus~\cite{Sirunyan:2020zal}.  At the next stage, 
a farm of processors
running a version of the full event reconstruction software optimized
for fast processing,
known as the high-level trigger (HLT), reduces the event rate to about
1\unit{kHz} before data storage~\cite{Khachatryan:2016bia}.
At the HLT, muon candidates are reconstructed in two steps. In the first step,
referred to as Level 2 (L2), muons are reconstructed using only the information
from the muon detectors, 
whereas at the stage referred to as Level 3 (L3), tracker information is also used.

\section{Signal models and simulated samples} \label{sec:samples}

\begin{figure}[htbp]
  \centering
  \includegraphics[width=0.80\DSquareWidth]{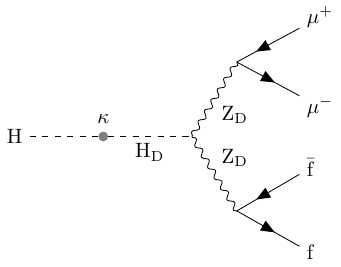}
\hspace*{3em}
  \includegraphics[width=0.45\linewidth]{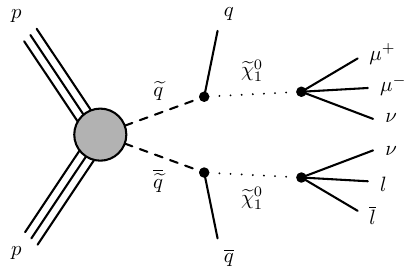}
   \caption{Feynman diagrams for (left) the HAHM model, showing the
    production of long-lived dark photons $\PZD$ via the Higgs portal,
    through $\PSMHiggs$--$\PDarkHiggs$ mixing with the parameter $\kappa$,
    with subsequent decays to pairs of muons or other fermions via the vector portal; and (right) pair 
    production of squarks followed by $\squarktoqchi$ decays, where the RPV neutralino is assumed to be a long-lived particle
    that decays into a neutrino and two charged leptons.
     \label{fig:SignalDiagram}}
\end{figure}

Two signal models with different final-state topologies and event
kinematics are used in the optimization of event selection criteria
and in the interpretation of results.  The first belongs to a class of
models featuring a ``hidden'' or ``dark'' sector of matter that does
not interact directly with the SM particles, but can manifest itself
through mixing effects.  This HAHM benchmark~\cite{Wells:2008xg,
  Curtin:2014cca} contains an extra dark gauge field
$U$(1)$_\text{D}$, which mixes kinetically with the hypercharge SM
gauge field (``vector portal'') and gives rise to a spin-1 mediator
known as the dark photon $\PZD$.  The symmetry of $U$(1)$_\text{D}$ is
broken by a new dark Higgs field $\PDarkHiggs$, which mixes with the
SM Higgs boson $\PSMHiggs$
(``Higgs portal'') and gives mass $\mZD$ to the dark photon.  If there are no
hidden-sector states with masses smaller than $\mZD$, the mixing
through the vector portal with the SM photon and \cPZ boson causes the
dark photon to decay exclusively to SM particles, with a sizable
branching fraction to leptons.  Pair production of the $\PZD$ via the
Higgs portal with subsequent decays of dark photons
via the vector portal is shown in Fig.~\ref{fig:SignalDiagram} (left).

The present search probes the regime of $\mZD > 10\GeV$ with small values of
the $\cPZ$--$\PZD$ kine\-tic mixing parameter $\epsilon$~\cite{Curtin:2014cca}.
In this regime, the dark photon is long-lived, since its mean proper lifetime
$\tau(\PZD)$ is proportional to $\epsilon^{-2}$.  In particular, the
dark photon with $10\GeV \lesssim \mZD < \mh/2$ is expected
to have macroscopically large mean proper decay lengths, for example, $\cTau(\PZD)
\gtrsim \mathcal{O}(100{\mum})$ for $\epsilon < \mathcal{O}(10^{-6})$.  The
$\PZD$ production rate is governed by the branching fraction 
$\mathcal{B}(\PSMHiggs \to \PZD\PZD)$, which does not depend on $\epsilon$ but is
proportional to the square of
$\kappa m^2(\PSMHiggs)/\abs{m^2(\PSMHiggs)-m^2(\PDarkHiggs)}$, where
$\kappa$ is the $\PSMHiggs$--$\PDarkHiggs$ mixing parameter.  Since
$\kappa$ and $m(\PDarkHiggs)$ affect only the overall dark photon
production rate, sampling of $\mZD$ and $\epsilon$ is sufficient to
explore different kinematical and topological scenarios of the model.
We generated a set of 24 simulated HAHM event samples with $\mZD$ between 10
and 60\GeV and $\epsilon$ between $10^{-7}$ and $2 \times 10^{-9}$.  In
this mass range, 
the model's prediction for $\mathcal{B}(\PZD \to \MuMu)$ varies
between 15.4\% at $\mZD = 10\GeV$ and 10.7\% at $\mZD = 60\GeV$.
The dark Higgs boson is assumed to be heavy enough so that
$\PSMHiggs \to \PDarkHiggs\PDarkHiggs$ decays are kinematically
forbidden. (In the sample generation, we use $m(\PDarkHiggs)=400\GeV$
and $\kappa=0.01$.)
The production of dark photons is modeled at leading order
by \MGvATNLO~\cite{Alwall:2014hca} version 2.9.9.
The samples are generated only for the dominant gluon-fusion
production mechanism, but the Higgs boson production cross section is
normalized to the most recent theoretical prediction for the sum of
all production modes for $\mh = 125\GeV$ at $\sqrt{s} = 13.6\TeV$,
59.8\unit{pb}~\cite{Cepeda:2019klc, LHCHWG136TeV}.  The decays of the dark photons
are modeled by \PYTHIA 8.306~\cite{Sjostrand:2014zea}.

We also consider a simplified benchmark model inspired by RPV SUSY
and featuring displaced $\MuMuNu$ vertices. Unlike the HAHM, where the two-body LLP decay leads
to displaced $\MuMu$ vertices that are reconstructed with the dimuon invariant mass $\mMuMu$ corresponding to the LLP mass,
the presence of a neutrino at the decay vertex leads to a nonpeaking $\mMuMu$ distribution with a
broad spectrum below the endpoint at the LLP mass.
In this model, which was used by CMS in the Run~1 searches for displaced 
dimuons~\cite{EXO-12-037, CMS-PAS-EXO-14-012} and by ATLAS in the Run~2 search for pairs of displaced charged leptons~\cite{Aad:2683781}, the LLP is assumed to be an RPV neutralino $\PRPVChi$ that results from decays of mass-degenerate squarks, \squarktoqchi, which
are pair produced in $\Pp\Pp$ collisions.
Nonzero values of RPV couplings $\lambda_{122}$ and $\lambda_{232}$ enable 
displaced $\PRPVChi$ decays into a pair of oppositely charged muons and a
neutrino, $\chitomumunu$~\cite{Dreiner:1997uz, Dercks:2017lfq}. The Feynman diagram
for this process is shown in Fig.~\ref{fig:SignalDiagram} (right). 

The search uses benchmark signal samples that are generated assuming ${\cal B}(\chitomumunu) = {\cal B}(\chitoeenu) = 0.5$, which
gives rise to events with up to two displaced dimuon vertices. 
To explore a wide range of kinematic variables and event topologies,
we chose six $\mSquark$ values in the range between 125\GeV and 1.6\TeV, and
for each chosen $\mSquark$, generated sets of samples with $\Delta m = \mSquark-\mChi$ of 25, 200, and 650\GeV,
and sets with constant $\mChi$ values of 50 and 500\GeV such that $\mChi< \mSquark$.
To study a wide range of signal displacements,
each set contains three samples with the generated $\cTau(\PRPVChi)$ values corresponding to mean transverse decay lengths of approximately 3, 30, and 250\cm
in the laboratory frame. 
All other SUSY particles (\eg, gluinos and sleptons) are assumed to be too heavy to be produced. (Their masses are set to 10\TeV.)
The samples are generated with \PYTHIA 8.306.  The squark-antisquark
production cross sections are calculated with {\tt NNLL-fast} version
2.0 to approximate next-to-next-to-leading order (NNLO) in the strong
coupling constant, including the resummation of soft gluon emission at
next-to-next-to-leading logarithmic accuracy~\cite{Beenakker:2016lwe}.
The computation uses the NNLO PDF4LHC21
parton distributions functions (PDFs)~\cite{PDF4LHCWorkingGroup:2022cjn}.

Since the optimization of the event selection criteria and the
evaluation of the residual backgrounds are performed using data, the
simulated background samples are used primarily to
gain a better understanding of the nature and composition of
background events passing the event selection.  Simulated background samples used in the analysis
include Drell--Yan (DY) dilepton production; $\ttbar$, $\PQt\PW$, and
$\PAQt\PW$ events; \PW and \PZ boson pair production (dibosons);
samples of \JPsi mesons produced in $\PQb$ hadron cascade decays; $\PW$+jets;
and events comprised of jets produced through the strong interaction that are
enriched in muons from semileptonic decays of hadrons containing $\PQb$ or $\PQc$
quarks.

The simulated signal and background samples are
produced with PDFs NNPDF3.1~\cite{Ball:2017nwa} at NNLO, 
using the CP5 tune~\cite{Sirunyan:2019dfx},
which is optimized for these PDFs, to model the underlying event.
The passage of particles through the detector 
is simulated by \GEANTfour~\cite{Agostinelli:2002hh}.
Simulated minimum bias events are superimposed on a hard interaction
in simulated events to describe the effect of additional inelastic
$\Pp\Pp$ interactions within the same or neighboring bunch crossings,
known as pileup.
All simulated events are then
reconstructed with the same algorithms as used for data.
A reweighting procedure that combines events in all simulated signal
samples at a given signal mass~\cite{Will_thesis} is employed to
calculate the efficiencies for lifetimes different from those of the
available samples.

\section{Analysis strategy and event selection} \label{sec:selection}

\subsection{Analysis strategy} \label{subsec:strategy}
An LLP produced in the hard interaction of the
colliding protons may travel a significant distance in the detector before
decaying into muons.  
While trajectories of the muons produced well within the silicon
tracker can be reconstructed by both the tracker and the muon system,
tracks of muons produced in the outer tracker layers or beyond
can only be reconstructed by the muon system.  Since the
dimuon vertex resolution and the background composition differ
dramatically depending on whether the muon is reconstructed in the
tracker, we classify all reconstructed dimuon events into three
mutually exclusive categories: a) both muons are reconstructed using
both the tracker and the muon system (TMS-TMS category); b) both muons
are reconstructed using only the muon system, as ``standalone'' muons
(STA-STA category); and c) one muon is reconstructed only in the muon
system, whereas the other muon is reconstructed using both the tracker
and the muon system (STA-TMS category).  These three categories of
events are analyzed separately, with each benefiting from dedicated event
selection criteria and background evaluation.
The STA-TMS category, which had the lowest sensitivity among the three
categories in the Run 2 analysis~\cite{EXO-21-006}, is not used for the results presented in this paper; the results in the other two categories are statistically
combined.

The beam spot is identified with the mean
position of the $\Pp\Pp$ interaction vertices.  
The primary vertex (PV) is taken to be the vertex corresponding to the
hardest scattering in the event, evaluated using tracking information
alone, as described in Section 9.4.1 of Ref.~\cite{CMS-TDR-15-02}.
A pair of reconstructed muon tracks is fitted to a common secondary vertex (SV), which is
expected to be displaced with respect to the PV.  
The transverse decay vector $\vecLxy$ is defined from the PV to the SV in the plane transverse to the beam direction, 
while the transverse decay length \Lxy is the magnitude of this vector.  
The transverse impact parameter $\dzero$ is defined as the distance of closest approach of
the muon track in the transverse plane with respect to the PV.

\subsection{Trigger algorithms} \label{subsec:trigger}
Events were collected with dedicated triggers aimed at recording dimuons produced both within and outside of the tracker.
In Run~2, these triggers required two muons reconstructed in the muon system alone, without using any information from the tracker~\cite{CMS:2021yvr}.
They were deployed in 2016 and 2018 and included a beam spot constraint in the muon track fits at \Lone but not at the HLT.
The values of the \Lone \pt thresholds varied from 11 and 4\GeV (for
the leading and subleading \Lone muons, respectively) during most of
2016 to 15 and 7\GeV at the end of Run~2.
Each HLT muon was required to be within the region $\abs{\eta} < 2.0$ and to have transverse momentum $\pt > 28~(23)\GeV$ for 2016 (2018) data taking.

The primary goal of the trigger optimization performed for Run~3 data taking was to increase the signal 
efficiency by lowering as much as possible the \pt thresholds and by removing the beam spot constraint at \Lone, without increasing considerably the resulting trigger rate.
Two additional sets of \Lone triggers were introduced.  One set
comprises double-muon triggers with either no or a very low (4.5\GeV) muon-\pt
threshold, which were made possible by requiring that the \Lone muon
candidates have opposite signs, segments in at least three different
muon stations, angular separation $\DeltaR_{\PGm\PGm} =
\sqrt{\smash[b]{(\Delta\eta_{\PGm\PGm})^2 +
    (\Delta\phi_{\PGm\PGm})^2}}$ that does not exceed a threshold that
varies between 1.2 and 1.4, and $\abs{\eta} < 1.5$ in the trigger with
no explicit \pt requirement.
The other set
takes advantage of a 
new track-finding procedure in the barrel section of the \Lone
muon trigger. This procedure enables the reconstruction of \Lone muon candidates 
and the determination of their \pt without using the beam spot constraint~\cite{CMS:2023gfb}.
These triggers require this \pt be larger than 15 and
7\GeV for the leading and subleading \Lone muon, respectively.  The \pt
thresholds are lowered to 6 and 4\GeV when $\dzero$ of both \Lone
muons is larger than 25\unit{cm}.
The new \Lone triggers mitigate the efficiency loss that was present in Run~2 for displaced muons
not pointing to the beamspot~\cite{EXO-21-006}, thereby improving the trigger efficiency for LLPs
with $\cTau \gtrsim 10\unit{cm}$.  The relative efficiency gain
evaluated using the simulated HAHM signal events increases as $\cTau$
increases, reaching a plateau at 20 to 50\% (depending on $\mZD$) at
$\cTau(\PZD) \approx 10\unit{m}$.

\begin{figure}
  \centering
  \includegraphics[width=\DFigWidth]{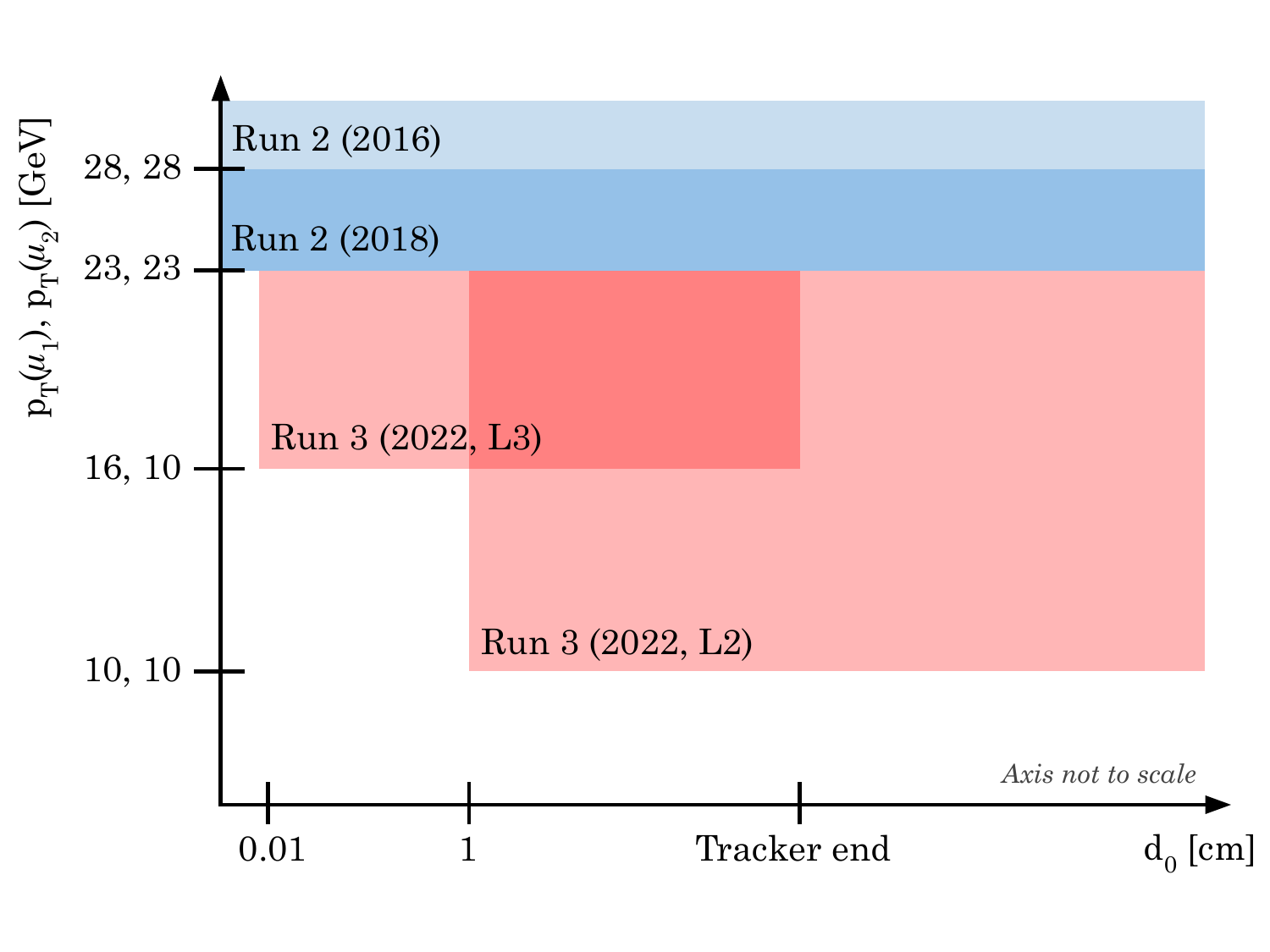}
  \caption{The \pt and \dzero coverage of the 2016 Run~2 triggers (light blue), 
  	2018 Run~2 triggers (blue), and newly designed 2022 Run~3 triggers described in the text (red).
  The two values of the \pt refer to the trigger thresholds for the
  muons.}
  \label{fig:run3_hlt_sketch}
\end{figure}

The events selected by the aforementioned \Lone triggers were then required to be selected by a logical OR of the 2018 L2 trigger paths used in the Run 2 analysis~\cite{EXO-21-006} and the newly designed HLT paths. The new paths use two complementary algorithms 
described below, and are characterized by \pt thresholds that depend on the muon \dzero, as illustrated in Fig.~\ref{fig:run3_hlt_sketch}.  
The first algorithm, labeled Run~3 (2022, L2) in Fig.~\ref{fig:run3_hlt_sketch}, 
imports some of the elements of the offline analysis into the online selection, in order to take advantage
of the superior tracker resolution in determining muon \dzero.
Instead of stopping the online reconstruction at the L2 stage (muon system alone) as it was done in the Run 2 trigger, the algorithm attempts to reconstruct the muon candidates
at the L3 stage as well (similar to offline TMS muons). If either of
the two L2 muon candidates is reconstructed at L3 as a muon with $\dzero < 1\cm$, the event is discarded, since such
an L2 muon candidate is likely to originate from the background processes.
The resulting trigger operates with muon \pt thresholds of 10\GeV
and improves significantly the signal efficiency in the STA-STA category, 
while contributing only about 1\% additional HLT rate. 

The second new HLT algorithm, labeled Run~3 (2022, L3) in Fig.~\ref{fig:run3_hlt_sketch}, introduces new paths relying
entirely on the online L3 muon reconstruction. A moderate threshold on the impact parameter of each muon, $\dzero > 0.01\cm$,
allows the \pt thresholds to be fairly low:
16\GeV on the leading muon $\PGm_1$ and 10\GeV on the subleading muon $\PGm_2$.
The resulting trigger greatly improves the signal efficiency in the TMS-TMS category, 
while adding only another $\approx$1\% of HLT rate.
An increase in the number of background events selected by the trigger
led to the refinements in the offline event selection described in
Section~\ref{subsec:recosel}.

\begin{figure}[htb]
  \centering
  \includegraphics[width=\DFigWidth]{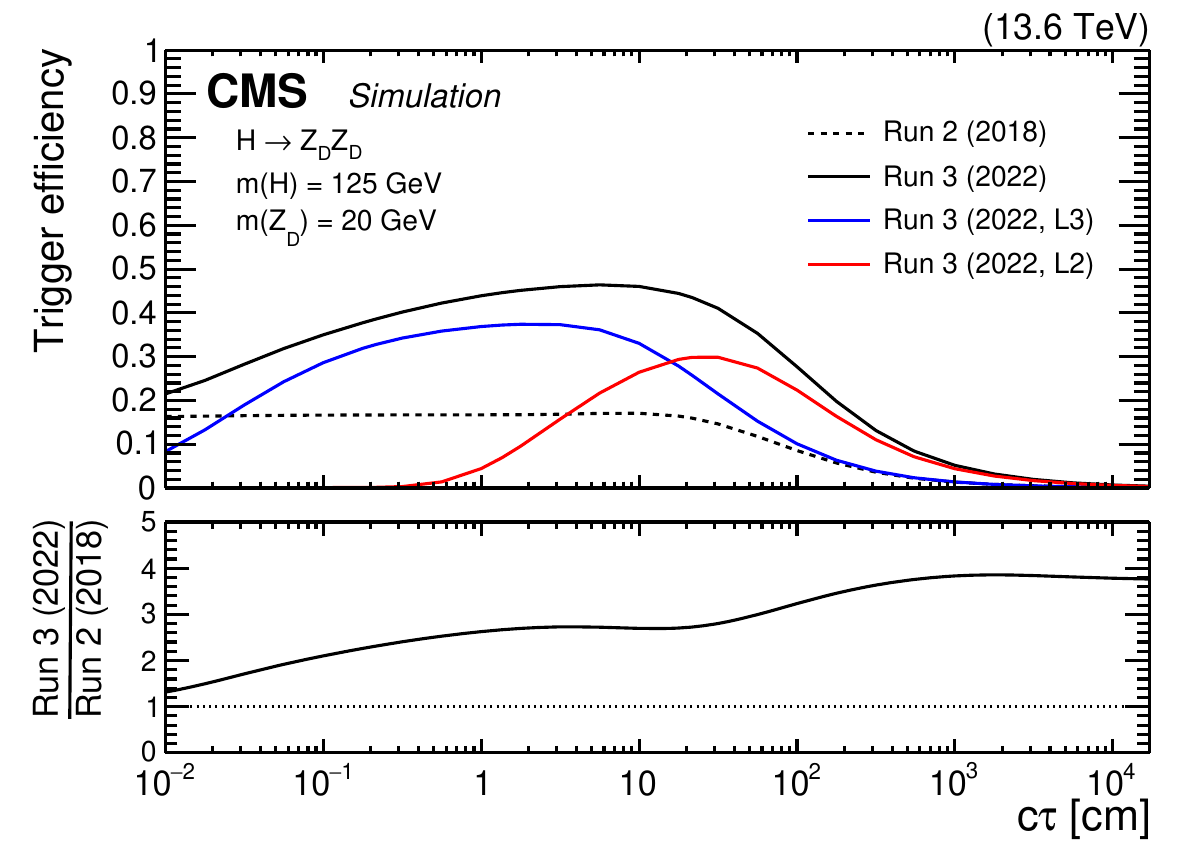}
  \caption{Efficiencies of the various displaced dimuon trigger paths and their combination as a function
    of $\cTau$ for the HAHM signal events with $\mZD = 20\GeV$.  The efficiency is defined as the fraction of
    simulated events that satisfy the detector acceptance and the requirements of the following
    sets of trigger paths: the Run~2 (2018) triggers (dashed black); the
    Run~3 (2022, L3) triggers (blue); the Run~3 (2022, L2) triggers
    (red); and the OR of all these triggers (Run~3 (2022), black).
    The lower panel shows the ratio of the
    overall Run~3 (2022) efficiency to the Run~2 (2018) efficiency.}
  \label{fig:HLT_efficiency_split_by_path}
\end{figure}

The combined L1-HLT efficiency of the various displaced dimuon
trigger paths and their combination as a function of \cTau is shown in
Fig.~\ref{fig:HLT_efficiency_split_by_path} for the simulated HAHM
signal events with $\mZD = 20\GeV$.
Because the Run~2 triggers  (dashed black) have no restrictions on \dzero, they
continue to have the highest efficiency (15\%) at $\cTau < 0.02\cm$.
The addition of the Run~3 (2022, L3) paths (blue) increases the overall efficiency (black) by more than a factor of 2 for $\cTau = 0.1$--1\cm.
The efficiency of this trigger starts to drop at $\cTau \gtrsim 5\cm$, when dimuons are 
produced beyond the innermost tracker layers and the L3 muon reconstruction efficiency decreases. 
At larger \cTau values, the addition of the Run~3 (2022, L2) paths (red) strongly contributes 
to the improvement of the signal efficiency, \eg, by more than a factor of 3 at $\cTau=1\unit{m}$. 
The decline of the efficiencies at the largest \cTau is driven by the increased fraction of dimuons produced outside the CMS detector. 
Overall, the addition of new \Lone and HLT paths improves the trigger efficiency for $\PZD$ with $\mZD > 10\GeV$ and $\cTau \gtrsim 0.1\cm$ by a factor of 2 to 4, depending on \cTau and mass.
The performance of the trigger was validated using data as described
in Section~\ref{sec:systunc_signal}.

\subsection{Muon reconstruction and event selection} \label{subsec:recosel}
Optimal performance for the wide range of displacements of secondary 
vertices considered in the analysis cannot be achieved by a single muon 
reconstruction algorithm. To accurately reconstruct muons produced near the IP, 
commonly used algorithms developed for prompt muons are employed. 
These algorithms combine measurements from both the tracker and the muon system.
Two such TMS algorithms are the global muon and tracker muon reconstruction 
algorithms~\cite{Chatrchyan:2012xi, Sirunyan:2018}. 
The global muon algorithm reconstructs muons by fitting hits 
in the tracker and segments in the muon system into a common track. 
The tracker muon algorithm, on the other hand, builds muons
by extrapolating tracks in the inner tracker to the muon system and 
requiring loose geometric matching to DT or CSC segments.
However, the efficiency of these algorithms decreases rapidly as the 
distance between the IP and the muon origin increases. 
In contrast, algorithms that rely solely on information 
from the muon system can still efficiently reconstruct muons produced in the
outer tracker layers and beyond. 
These STA algorithms~\cite{Chatrchyan:2012xi, Sirunyan:2018}
can reconstruct muons with displacements of up to a few meters. 
However, they exhibit poorer spatial and momentum resolution compared to muons 
reconstructed using the more precise information from the silicon tracker.

To benefit from the advantages offered by both types of algorithms, 
we begin the muon selection with the muons reconstructed by a specific STA algorithm 
that eliminates the beam spot constraints from all stages of the muon reconstruction procedure. 
This approach yields the highest efficiency and the 
finest resolution for highly displaced muons, surpassing all other available STA algorithms.
Subsequently,
we attempt to match each STA muon with muons
reconstructed using global muon and tracker muon algorithms,
and replace the STA muon with an associated TMS muon if a match is found.
As in the Run~2 analysis~\cite{EXO-21-006}, we reject events in which
no HLT muon pair that triggered the event matches two STA muons.

Due to the need to reduce large backgrounds as much as possible, 
careful optimization of the event, muon, and dimuon selection was done for the Run~2 analysis, as described in Section~4 of Ref.~\cite{EXO-21-006}.
A summary of the selection criteria used in the Run~2 analysis is
given in Table~1 of Ref.~\cite{EXO-21-006}.
Most of the selection criteria are unchanged,
as is the association between STA and TMS muons for
those tracks originating within the tracker volume.
The rest of this section of the paper describes only the changes 
and refinements to the muon and dimuon selection for this Run~3 search, while also introducing analysis
variables that are used in the background estimation. 

Because of the increase in the background as a result of the lower \pt trigger thresholds,
an isolation requirement is useful when applied to the STA muons. Both STA muons in the STA-STA category are
required to have relative tracker isolation $\tkisoSTA<0.15$, where $\tkisoSTA$ is defined as the
sum of the \pt of tracks within a cone of $\DeltaR = \sqrt{\smash[b]{(\Delta\eta)^2 + (\Delta\phi)^2}} < 0.3$ around the
trajectory of the muon extrapolated to the point of closest approach
to the PV divided by the muon track \pt.
When evaluated using events that pass all other selection criteria,
this isolation requirement is more than 90\% efficient for signal and 
suppresses background from events with jets that are produced through the strong interaction,
collectively referred to as quantum chromodynamics (QCD) events,
by about a factor of 2.
The TMS muon isolation criterion, $\tkisoTMS < 0.075$, remained
unchanged with respect to the Run~2 analysis~\cite{EXO-21-006}.

The Run~2 analysis required that the dimuons be displaced with respect to the
PV by imposing requirements on the \Lxy significance \LxySig, where \LxyErr is the uncertainty in $\Lxy$~\cite{EXO-21-006}. 
A significant fraction of remaining events in the STA-STA category
suffer from unreliable SV reconstruction; we thus require in addition that $\LxyErr<20\cm$.

In the TMS-TMS category, the sensitivity of the search is further improved by binning events in min($\dzeroSig$),
which is the minimum value of the ratio of $\dzero$ to its uncertainty for the two muons forming a dimuon.   
Unlike signal events with macroscopic \cTau values, the expected background 
has a steeply falling min($\dzeroSig$) distribution, which motivates the splitting 
of the signal region (SR) into three bins of min($\dzeroSig$) ranges. 
The bin ranges are the same as those in the Run~2 analysis, namely 6--10, 10--20,
and ${>}20$.

\begin{figure}
  \centering
  \includegraphics[width=0.95\DSquareWidth]{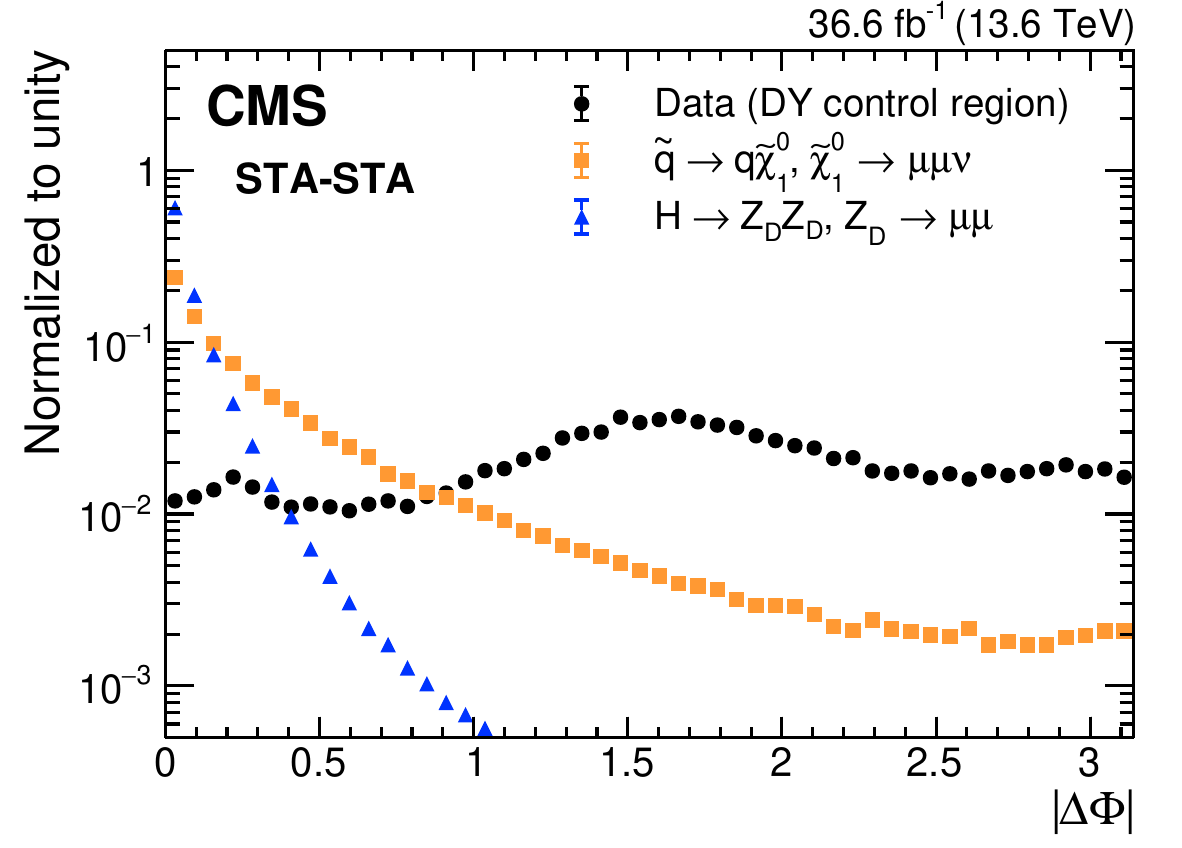}
  \includegraphics[width=0.95\DSquareWidth]{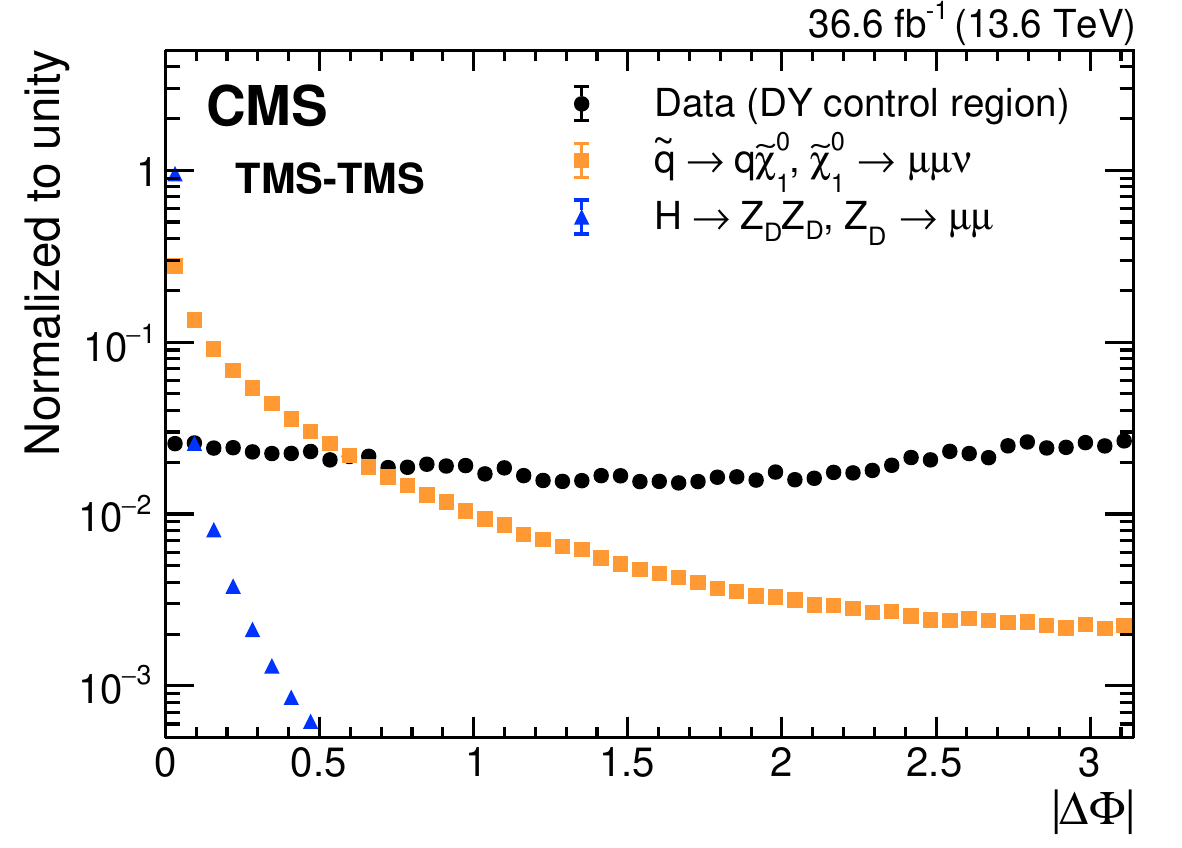}
  \caption{Distributions of \DeltaPhiAbs for (left) STA-STA and
    (right) TMS-TMS dimuons in data samples obtained by inverting some
    of the selection criteria and enriched in DY events (black
    circles) and for events passing all selection criteria except for
    a requirement on \DeltaPhiAbs in all HAHM (blue triangles) and RPV
    SUSY (orange squares) generated signal samples combined.  All
    distributions are normalized to unit area.}
  \label{fig:deltaPhi_Sig}
\end{figure}

A quantity that is useful for suppressing and evaluating the SM background
that satisfies all other selection requirements is the azimuthal angle
$\DeltaPhi$ between $\vecLxy$ and the transverse momentum vector $\vecptMuMu$ of the dimuon system.
A major source of SM background events is from prompt high-mass dimuons
that are reconstructed as displaced due to instrumental or
reconstruction failures.  Such dimuons mostly arise from DY
dimuon production.  Events from DY $\PGtp\PGtm$ production with both
$\Pgt$ leptons decaying to muons lead to a background with
characteristics similar to those of the mismeasured DY
$\MuMu$ events; contributions from processes such as $\ttbar$ and
diboson production are relatively small.  
Dimuons originating in 
these events, collectively referred to as DY events, are expected to have a distribution of
\DeltaPhiAbs symmetric about $\pi/2$, because the 
dimuon momentum vector is uncorrelated with the \vecLxy vector.
On the other hand, when a pair of muons is produced in the decay of an LLP originating at the PV, the resulting
$\vecptMuMu$ and $\vecLxy$ are collinear, and the distribution of \DeltaPhiAbs peaks at zero.  This can be seen in 
Fig.~\ref{fig:deltaPhi_Sig}, which compares the \DeltaPhiAbs distributions of TMS-TMS and STA-STA
dimuons in data samples obtained by inverting some of the selection
criteria and enriched in DY events with the \DeltaPhiAbs distributions
for events passing the full selection except for the \DeltaPhiAbs
requirement in all HAHM and RPV SUSY signal samples combined.
There is a small asymmetry in the \DeltaPhiAbs distribution for DY events in the STA-STA category, which is caused by the event selection criteria.

To address the different $\DeltaPhiAbs$ distributions in the two types of signal models, 
we define different critical values $\DeltaPhiAbsC$ for requiring $\DeltaPhiAbs<\DeltaPhiAbsC$.
To define the SR for the HAHM model, where $\DeltaPhiAbs$ strongly peaks at zero and the tail is driven by the resolution, 
we use $\DeltaPhiAbsC = \pi/10$ 
in the STA-STA category and $\pi/30$ in the TMS-TMS category.
The tighter requirement in the TMS-TMS category takes advantage of the better tracker resolution. 
For the RPV SUSY model, where the distribution is broader due to the undetected neutrino among the decay products, 
we use a looser requirement of the Run 2 analysis, $\DeltaPhiAbsC = \pi/4$, in both dimuon categories. 
For the chosen $\DeltaPhiAbsC$ values, the signal efficiency is 90--99\% in the HAHM  
and 60--99\% in the RPV SUSY models, depending on the Lorentz boost of the LLP;
the corresponding DY background rejection factors are, respectively,
15--20 and 4--6, depending on the dimuon category.
We use the symmetric region, $\DeltaPhiAbs > \pi -\DeltaPhiAbsC$,
as a control region (CR) for evaluating the contribution from
DY and other prompt backgrounds, and the validation regions (VR) with $\pi/4 < \DeltaPhiAbs < \pi/2$ 
and $\pi/2 < \DeltaPhiAbs < 3\pi/4$ for validating background predictions, as discussed in 
Section~\ref{sec:bkg_evaluation}.

In addition to defining \DeltaPhiAbs SR and CRs, we
classify selected dimuons as oppo\-site-sign (OS) or same-sign (SS), based on the observed muon charges.   The signal selection requires that dimuons be OS, while SS dimuons constitute a CR
used to evaluate backgrounds arising from QCD events.

The last important source of SM backgrounds consists of QCD events in which there are
di\-muons produced in decays of \JPsi mesons and other 
low-mass SM resonances, or formed from the products of the $\PQb$ hadron 
cascade decays ($\PQb \to \PQc\PGm_{1}X$ followed by $\PQc
\to \PGm_{2}X$).  These events are suppressed by requiring that $\mMuMu > 10\GeV$.
However, it was observed that low-\pt muons
can appear as muons with higher \pt, with straighter tracks, when
reconstructed from a small number of measurements.
This gives rise to dimuons with an overestimated $\mMuMu$ (above the
10\GeV threshold) and a mistakenly formed displaced vertex. 
These dimuons typically have small $\DeltaPhiAbs$ 
(either due to the collinearity of the $\vecptMuMu$ and \vecLxy vectors
or overestimated muon \pt) and may exhibit 
large values of \LxySig and $\dzeroSig$ that resemble signal-like characteristics. 
To suppress such events, which are particularly abundant in the 
STA-STA category due to its low resolution, we reject STA-STA dimuons whose separation
in $\eta$ is small ($\DetaMuMu < 0.1$)
if one of the muons is reconstructed in the barrel from fewer than 25
DT hits or if the sum of the segments belonging to both muons in the dimuon
is fewer than 6.  These requirements
are identical to those used in the Run~2 analysis.

Finally, to test for the existence of an LLP with a given mass,
dimuons satisfying the selection criteria are required to have
\mMuMu within a specified interval 
containing the probed LLP mass.  The width of each interval is chosen
according to the mass resolution and the expected background.  
For the interpretation in the framework of the HAHM model, where the LLP decay
products do not contain any undetected particles, the full widths of the chosen \mMuMu
intervals are approximately equal to 6--8 times the mass
resolution at this mass.  This choice typically yields intervals
containing a large fraction (90--99\%) of
putative signal with the probed mass.  Since the mass resolution in
the TMS-TMS category is far superior to that in the STA-STA
category (1--3\% compared to 10--25\%, for LLP masses between 20 and
350\GeV), the minimum width of \mMuMu intervals varies from 3\GeV in the
TMS-TMS category to ${\approx}20\GeV$ in the STA-STA category.

In the RPV SUSY model, the presence of a neutrino in the $\chitomumunu$ decay vertex 
leads to a nonpeaking dimuon invariant mass distribution with an endpoint 
at $\mChi$. To improve the mass resolution and the signal-to-background
discrimination in signals featuring an LLP decay vertex formed by a dimuon plus neutral 
or unobserved particles, we use the quantity referred to as the corrected SV mass, defined as
\begin{linenomath}
\begin{equation}
\mMuMuCorr = \sqrt{\mMuMu^2 + p_{\PGm\PGm}^2\sin^{2}{\theta}} + p_{\PGm\PGm}\sin{\theta}, 
\label{eq:corrMass}  
\end{equation} 
\end{linenomath}
where $p_{\PGm\PGm}$ is the magnitude of the dimuon momentum vector $\vec p_{\PGm\PGm}$, and $\theta$ is 
the angle between $\vec p_{\PGm\PGm}$ and the vector
connecting the PV with the SV.
The motivation behind this variable,
which was used at LEP and the SLC and recently by
LHCb~\cite{Aaij:2012990} and CMS~\cite{Sirunyan:2298594},
is to rely on the conservation of 
momentum and correct for the missing mass of the vertex due to unaccounted particles. 
In the case where only the dimuon belongs to the LLP decay vertex (and before measurement error), $\theta=0$ and $\mMuMuCorr = \mMuMu$. 
If, however, the decay vertex includes other particles, then $\theta>0$, $\mMuMuCorr > \mMuMu$, and the corrected mass is the 
minimum mass that the LLP, in this case the $\PRPVChi$, can have in order to be consistent 
with the measured direction of flight.
The $\mMuMuCorr$ distributions of simulated RPV SUSY events
show a clear peak at $\mChi$ in both dimuon categories.
We profit from the improvement in mass resolution offered by \mMuMuCorr and use \mMuMuCorr intervals to probe different $\mChi$.
As with $\mMuMu$ intervals for the HAHM model, the width of the intervals
is chosen to contain a large fraction of the RPV SUSY signal (typically 80--90\%).
In the STA-STA category, we additionally require
$\mMuMu > 15\GeV$ to avoid the leakage of background events from small
$\mMuMu$ to large $\mMuMuCorr$.

\begin{figure}[htb]
  \centering
  \includegraphics[width=\DFigWidth]{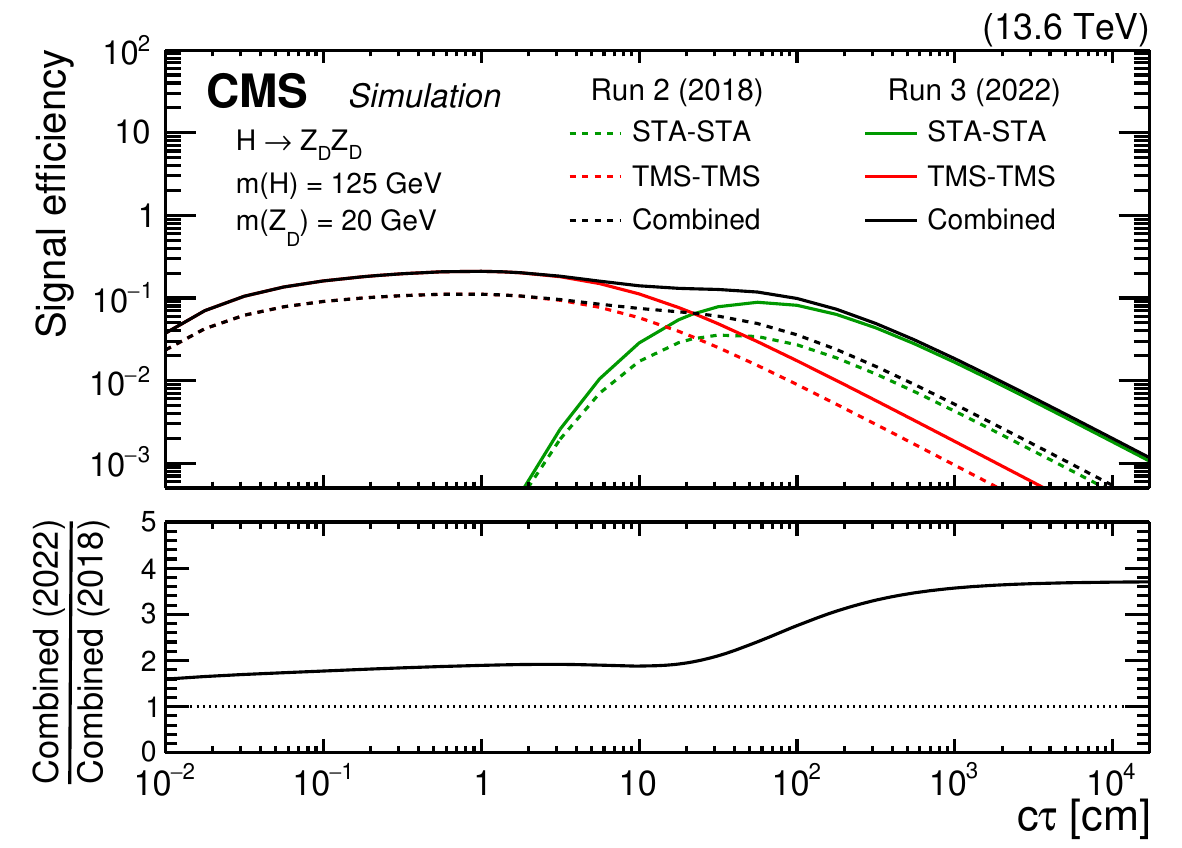}
  \caption{Overall efficiencies in the STA-STA (green) and 
    TMS-TMS (red) dimuon categories, as well as their combination (black)
    as a function of $\cTau$  for the HAHM signal events with $\mZD=20\GeV$.
    The solid curves show efficiencies achieved with the 2022 Run 3 triggers,
    whereas dashed curves show efficiencies for the subset of events
    selected by the triggers used in the 2018 Run 2 analysis.
    The efficiency is defined as the fraction of signal events that satisfy the criteria of 
    the indicated trigger as well as the full set of offline selection criteria.
    The lower panel shows the relative improvement of the overall signal efficiency brought in by improvements in the trigger.}
  \label{fig:overall_eff_Zd_ctau}
\end{figure}

Figure~\ref{fig:overall_eff_Zd_ctau} shows, as an example, the full
event selection
efficiency for the HAHM signal with $\mZD=20\GeV$ as a function of
\cTau, and illustrates relative improvements in the efficiencies
from the new trigger algorithms.  A large
increase in the overall efficiency, shown in the lower panel, is
apparent; most of the efficiency gain from the improvements in
the trigger algorithms and shown in
Fig.~\ref{fig:HLT_efficiency_split_by_path} is retained after the full
offline event selection.  Similar efficiency increases are seen at
other probed $\PZD$ masses.

\section{Background estimates and their systematic uncertainties} \label{sec:bkg_evaluation}

Background events that meet the event selection criteria cannot be reliably simulated 
since they consist of misreconstructed prompt muons and muons in jets. Therefore, 
we rely on analyzing events in the recorded data to estimate the expected background. 
To achieve this, we employ CRs where one or more selection criteria are inverted, 
creating a region predominantly populated by a specific type of background while having an 
insignificant contribution from the signal processes. The specific definitions of these 
CRs and the procedure for evaluating the background differ across dimuon 
categories and are elaborated on in the rest of this section.
The background evaluation methods are nearly all identical to those employed in
the Run 2 analysis~\cite{EXO-21-006}.  The only exception
is the evaluation of QCD backgrounds in the STA-STA category, which
can now benefit from events rejected by the new STA muon isolation requirement.
In order to prevent 
potential bias in the event selection, the events that satisfy the full set of selection 
criteria (\ie, those in the SR) were kept concealed ("blinded") until the final stages of the analysis.

One particular type of background events arises from
cosmic ray muons crossing the detector within the acceptance of the muon
system.  Such muons are often reconstructed as two back-to-back muons, one in the
upper half and one in the lower half of the detector.
For each dimuon category, the contribution from cosmic ray muons is 
assessed independently by examining the number of dimuons that meet 
all the selection criteria except for angular requirements designed to reject putative dimuons that are formed from back-to-back muons~\cite{EXO-21-006}. 
To evaluate this contribution, we use the rejection factors of these
requirements, which are determined from a sample of cosmic ray muons recorded 
during periods without any beam activity. In both dimuon categories, the 
remaining background originating from cosmic ray muons is estimated to be less 
than 0.1 events in all mass intervals combined.

\subsection{Estimation of Drell--Yan and other prompt backgrounds} \label{subsec:bkg_evaluation_DY}

The contribution from misreconstructed prompt high-mass
dimuons, mainly originating from the DY process, 
to the total background is evaluated from events in the signal-free
$\DeltaPhiAbs$-symmetric CR as
\begin{linenomath}
\begin{equation}
   N_{\text{DY}}^{i}(\text{OS}; \DeltaPhiAbs < \DeltaPhiAbsC) =
   N_{\text{DY}}^{i}(\text{OS}; \DeltaPhiAbs > \pi - \DeltaPhiAbsC) 
   R_\text{DY}^{i}{},
  \label{eq:DYestimate}
\end{equation}
\end{linenomath}
where $N_{\text{DY}}^{\text{i}}(\text{OS}; \DeltaPhiAbs < \DeltaPhiAbsC)$ and
$N_{\text{DY}}^{\text{i}}(\text{OS}; \DeltaPhiAbs > \pi - \DeltaPhiAbsC)$ are,
respectively, the numbers of DY background events in the
SR and its $\DeltaPhiAbs$-symmetric CR with $\DeltaPhiAbsC$ representing the selection criterion
applied;
$R_\text{DY}^{i}$ is the transfer factor accounting for the residual
asymmetry in the population of events in the two $\DeltaPhiAbs$ regions
and obtained from auxiliary measurements; and
the index $i$ denotes the dimuon category (STA-STA or
TMS-TMS).  The number of DY dimuons in the CR is obtained from the total number of events in that
region corrected by the expected contribution from QCD background
events, estimated as discussed in
Section~\ref{subsec:bkg_evaluation_QCD}.

To assess the symmetry of the $\DeltaPhiAbs$ distributions 
in this class of background events, events in dedicated CRs 
are used. In the STA-STA category, we focus on events 
within data CRs that are obtained by reversing the STA-to-TMS 
association. Specifically, we select events where STA-STA dimuons 
satisfy all the selection criteria, except that each of the constituent STA muons is associated with a TMS muon.
To ensure that these STA-STA dimuons are promptly produced 
(and not part of the signal), we require that the associated TMS-TMS dimuons, 
which offer superior spatial resolution, are prompt. 
This is achieved by imposing $\LxySig<1.0$ for the 
associated TMS-TMS dimuon.
To minimize contamination from muons originating from jets (which are discussed separately below), 
each TMS muon in the associated TMS-TMS dimuon is required to have
$\tkisoTMS < 0.05$.
From this sample, we obtain
$R^\text{STA-STA}_\text{DY} = 0.75 \pm 0.02$ (stat.) for $\DeltaPhiAbsC = \pi/10 $ and $R^\text{STA-STA}_\text{DY} = 0.73 \pm 0.02$ (stat.) for $\DeltaPhiAbsC = \pi/4$. The value of $R^\text{TMS-TMS}_\text{DY}$ is set to 
unity based on studies using simulation and data~\cite{EXO-21-006}.

In both STA-STA and TMS-TMS categories, no significant dependence of the transfer
factor $R_\text{DY}^{i}$ on reconstructed dimuon mass is observed, and a
single value is used for all signal mass intervals.  
The systematic uncertainties in $R_\text{DY}^\text{STA-STA}$ are assessed by comparing
$R_\text{DY}^\text{STA-STA}$ measured in individual mass intervals with the
result of the inclusive measurement and by varying
the boundaries and definitions of the auxiliary CRs.
The latter includes
repeating the measurements of $R_\text{DY}^{\text{STA-STA}}$ in the
region with only one STA-to-TMS association.
Based on these studies, we assign a systematic uncertainty of 30\%  in
$R_\text{DY}^{\text{STA-STA}}$.
The systematic uncertainty in $R_\text{DY}^{\text{TMS-TMS}}$ is assigned to
be 15\% based on the studies of the residual \DeltaPhiAbs asymmetry for
events in the min($\dzeroSig$) and \DeltaPhiAbs side bins.

\subsection{Estimation of QCD backgrounds} \label{subsec:bkg_evaluation_QCD}

Many of the background processes yielding small-$\DeltaPhiAbs$ OS
dimuons also give rise to small-$\DeltaPhiAbs$ SS dimuons, either
because these processes are charge symmetric or via muon charge
misassignment.  Most of these dimuons are embedded in jets and are
suppressed by the muon isolation requirements.
Thus, we evaluate the contribution from the QCD backgrounds to the SR from
events in CRs obtained by inverting two independent 
selection requirements: the isolation requirement (\ie, we select events
where at least one of the muons fails to satisfy the isolation criteria used in the analysis)
and the requirement of the OS dimuons (\ie, we select SS dimuons). 

In the STA-STA category, we base our estimate of the QCD backgrounds on the number of
events with OS nonisolated dimuons, using
\begin{linenomath}
\begin{equation}
  N_{\text{QCD}}^{\text{STA-STA}}(\text{OS}; \DeltaPhiAbs < \DeltaPhiAbsC) =
  N_\text{noniso}^{\text{STA-STA}}(\text{OS}; \DeltaPhiAbs < \DeltaPhiAbsC)
  R_\text{QCD}^{\text{STA-STA}}{},
 \label{eq:QCDestimate_DSADSA}
\end{equation}
\end{linenomath}
where $N_{\text{QCD}}^{\text{STA-STA}}(\text{OS}; \DeltaPhiAbs < \DeltaPhiAbsC)$
and $N_\text{noniso}^{\text{STA-STA}}(\text{OS}; \DeltaPhiAbs < \DeltaPhiAbsC)$
are, respectively, the numbers of OS isolated and nonisolated dimuons with small $\DeltaPhiAbs$, and
$R_{\text{QCD}}^{\text{STA-STA}}$ is the transfer factor between these numbers.
The transfer factor is evaluated using SS dimuons in the same small-$\DeltaPhiAbs$ region:
\begin{linenomath}
\begin{equation}
R_\text{QCD}^{\text{STA-STA}} =
\frac{N^\text{STA-STA}_{\text{iso}}(\text{SS}; \DeltaPhiAbs <
\DeltaPhiAbsC)}{N^\text{STA-STA}_{\text{noniso}}(\text{SS}; \DeltaPhiAbs <
\DeltaPhiAbsC)}{},
 \label{eq:QCDtransferfactor_DSADSA}
\end{equation}
\end{linenomath}
with $N^\text{STA-STA}_{\text{iso}}(\text{SS}; \DeltaPhiAbs < \DeltaPhiAbsC)$ and
$N^\text{STA-STA}_{\text{noniso}}(\text{SS}; \DeltaPhiAbs < \DeltaPhiAbsC) $ representing the
numbers of, respectively, isolated and nonisolated SS STA-STA dimuons
with $\DeltaPhiAbs < \DeltaPhiAbsC$. To improve the precision of the
$R_{\text{QCD}}^{\text{STA-STA}}$ measurement, we enlarge the sample of SS dimuons 
by removing tight requirements on the numbers of DT hits and muon
segments applied to dimuons with $\DetaMuMu < 0.1$ (discussed in
Section~\ref{sec:selection}) as well as the generic requirement on the
minimum number of DT hits~\cite{EXO-21-006}.  We obtain
$R_{\text{QCD}}^{\text{STA-STA}}$ values varying between 1.0 and 1.5
as a function of reconstructed dimuon mass and measured with an
uncertainty of 20--40\%.  They are in good agreement with
$R_{\text{QCD}}^{\text{STA-STA}}$ values obtained from a sample of OS
small-$\DeltaPhiAbs$ events failing the criteria for dimuons with
$\DetaMuMu < 0.1$.
Systematic uncertainties arising from these two measurements are found to be much smaller than statistical uncertainties resulting from CR sample sizes.

Based on the Run~2 studies~\cite{EXO-21-006}, we do not anticipate a significant 
contribution from $\DeltaPhiAbs$-asymmetric low-mass dimuons in the TMS-TMS category. 
This is primarily due to the superior dimuon invariant mass resolution in this category. 
However, since there may still be a small contribution from $\DeltaPhiAbs$-asymmetric dimuons,
we prefer not to rely solely on 
the symmetry of $\DeltaPhiAbs$ in evaluating the background. 
Consequently, the method used is similar to that employed in the
STA-STA category and bases our estimate of the QCD backgrounds on the
number of SS dimuons, using
\begin{linenomath}
\begin{equation}
  N_{\text{QCD}}^{\text{TMS-TMS}}(\text{OS}; \DeltaPhiAbs < \DeltaPhiAbsC) =
  N^{\text{TMS-TMS}}(\text{SS}; \DeltaPhiAbs < \DeltaPhiAbsC)
  R_\text{QCD}^{\text{TMS-TMS}}{}.
 \label{eq:QCDestimate_TMSTMS}
\end{equation}
\end{linenomath}
The transfer factor $R_{\text{QCD}}^{\text{TMS-TMS}}$ is obtained from
the ratio of OS to SS dimuons in the CR with the muon
isolation requirement reversed, which consists of dimuons passing the
nominal event selection but with at least one muon with $\tkisoTMS >
0.075$ and both with $\tkisoTMS < 0.5$.  We have verified that these
events, as well as SS dimuons passing isolation requirements, contain
negligible contributions from signal and DY events.
As the SR is divided into several min($\dzeroSig$) bins, the
evaluation of $R_{\text{QCD}}^{\text{TMS-TMS}}$ is performed
separately in each min($\dzeroSig$) bin.  Since no significant dependence of the
value of $R_{\text{QCD}}^{\text{TMS-TMS}}$ on $\mMuMu$ and $\mMuMuCorr$ is observed,
$R_{\text{QCD}}^{\text{TMS-TMS}}$ in each min($\dzeroSig$) bin is
calculated by integrating events in the entire invariant mass spectrum.
The measured values of $R_{\text{QCD}}^{\text{TMS-TMS}}$ decrease from
2.0 to 1.3 as min($\dzeroSig$) increases, with the statistical
uncertainties in the range 5--15\%.
A systematic uncertainty of 15\% is assigned to account for variations of
$R_{\text{QCD}}^{\text{TMS-TMS}}$ as a function of the invariant mass
and as the result of varying the definition and boundaries of the
auxiliary CR.

To avoid potential overestimation of the DY backgrounds, 
we use the same QCD background evaluation technique for dimuons 
in the CR defined by $\DeltaPhiAbs > \pi - \DeltaPhiAbsC$. 
The obtained
estimate of the QCD backgrounds is subtracted from the total to derive
the estimate of DY dimuons in this $\DeltaPhiAbs$ region, which is used for the
evaluation of the DY backgrounds in the SR according to Eq.~(\ref{eq:DYestimate}).  
This procedure
is not applied in the STA-STA category, where the
$\DeltaPhiAbs$-symmetric QCD background is negligible.  The sum of the
QCD and DY background estimates constitutes the total predicted
background in the SR.  According to the background
evaluation method, the DY backgrounds are
expected to dominate at small $\dzeroSig$ and $\LxySig$ values,
whereas the relative QCD contribution becomes larger as $\dzeroSig$
and $\LxySig$ increase.  The uncertainty in the background predictions 
is mainly driven by the statistical uncertainty arising from the 
limited number of events observed in the CRs.

\subsection{Validation of background predictions} \label{subsec:bkg_validation}

To ensure the reliability of the background evaluation method described 
in Sections~\ref{subsec:bkg_evaluation_DY} and \ref{subsec:bkg_evaluation_QCD},  
we examine the method's performance in various VRs, 
which are chosen to have minimal or no contribution from the signal being studied. 
Thus, we can verify that it accurately predicts 
the background and that any observed discrepancies are within acceptable limits.

\begin{figure}
  \centering
\includegraphics[width=0.97\DSquareWidth]{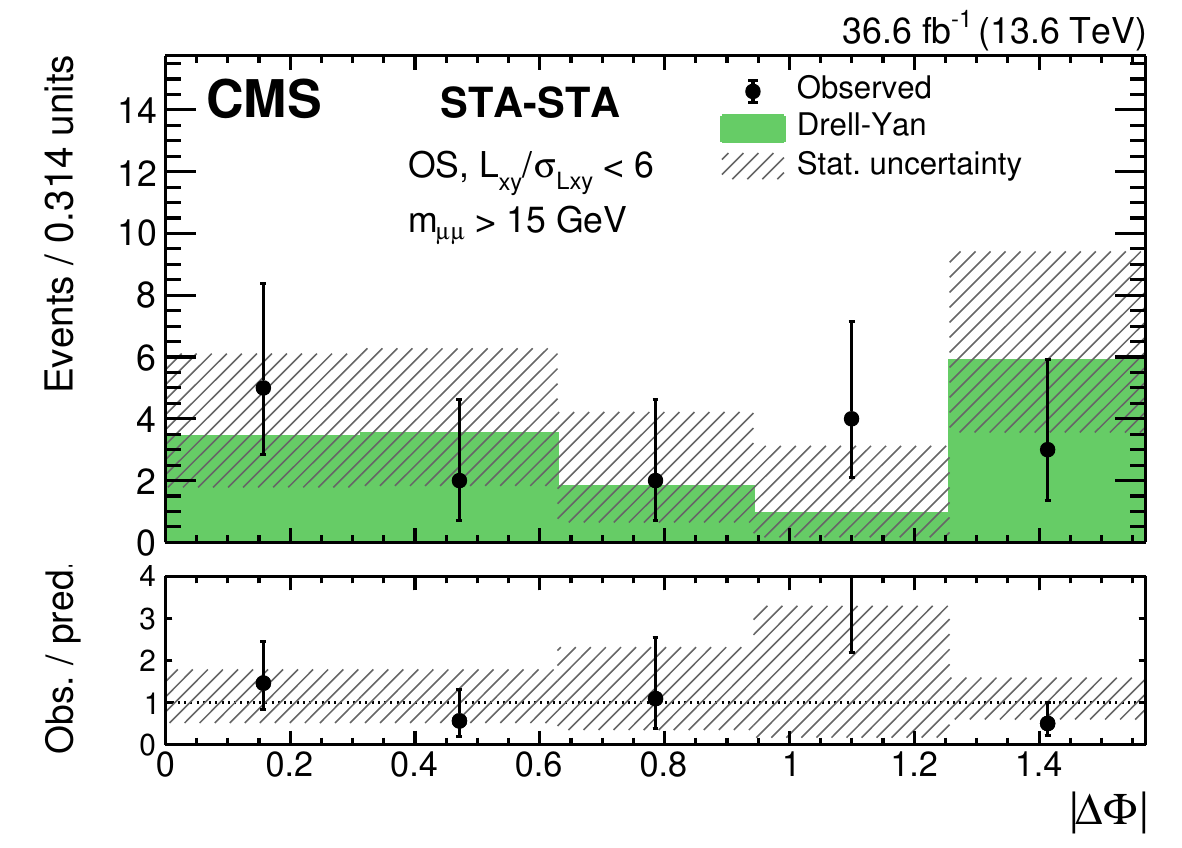}
\includegraphics[width=0.97\DSquareWidth]{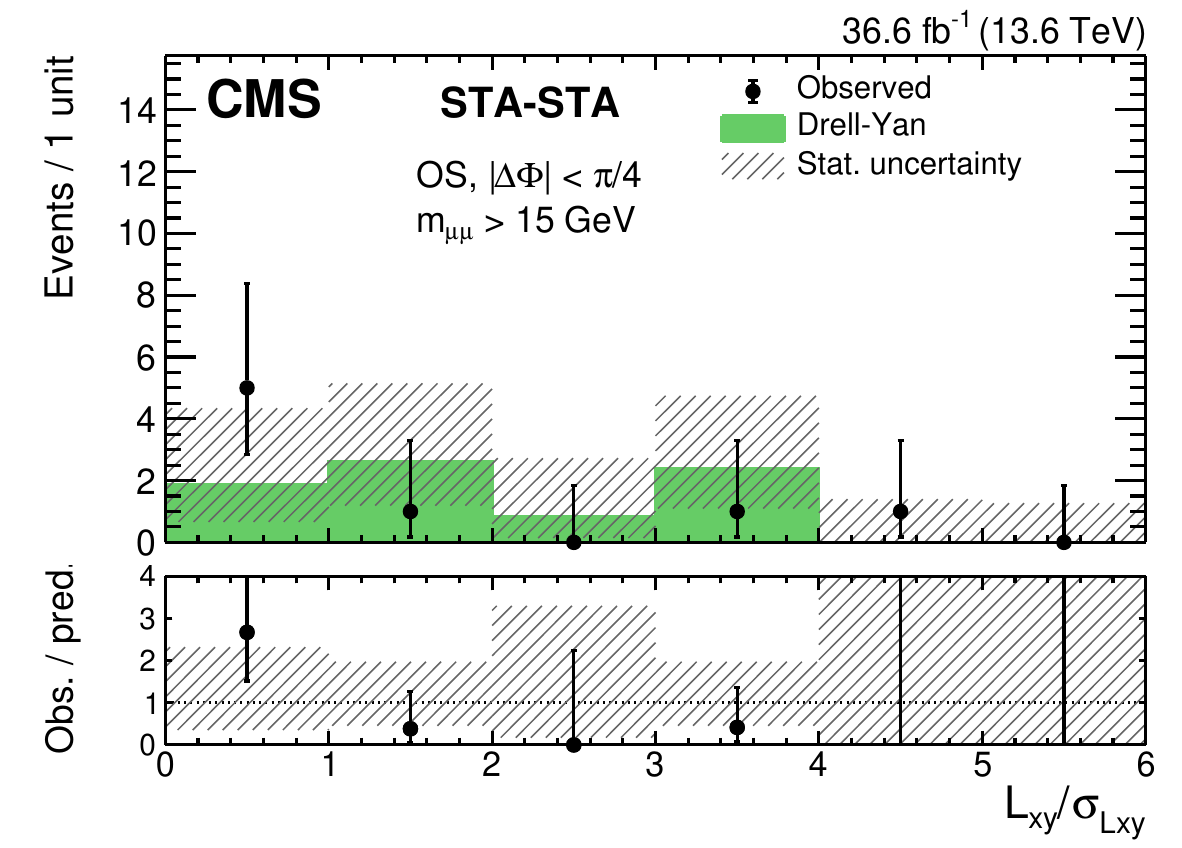}
  \caption{Example of background prediction checks in the STA-STA category: 
  distributions of (left) $\DeltaPhiAbs$ and (right) $\LxySig$ for events
  with $\mMuMu > 15\GeV$ in the $\LxySig < 6$ validation region in data (black circles)
  	 compared to the background predictions (histograms). 
	 The lower panels show the ratio of the observed to predicted number of events.  
	 Hatched histograms show the statistical uncertainty in the background prediction.}
    \label{fig:DY_closure}
\end{figure}

The evaluation of DY backgrounds is
examined in the VRs obtained by inverting the $\LxySig$ and
$\dzeroSig$ requirements and thereby enriched in this class of events.  
An example is shown in
Fig.~\ref{fig:DY_closure}, which compares the background
predictions to the observed distributions in the $\LxySig < 6$
VR for dimuons with $\mMuMu > 15\GeV$ in the STA-STA category.  
To check how the validity of the method depends on $\DeltaPhiAbs$ and the dimuon
displacement, we perform the evaluation of the background in corresponding
bins. The expected number of background events in the
given VR is computed according to
Eq.~(\ref{eq:DYestimate}) separately
in each $\DeltaPhiAbs$ or $\LxySig$ bin and compared to the observed data.  
The predictions of the method are consistent with the yields in data.

\begin{figure}
  \centering
  \includegraphics[width=0.95\DSquareWidth]{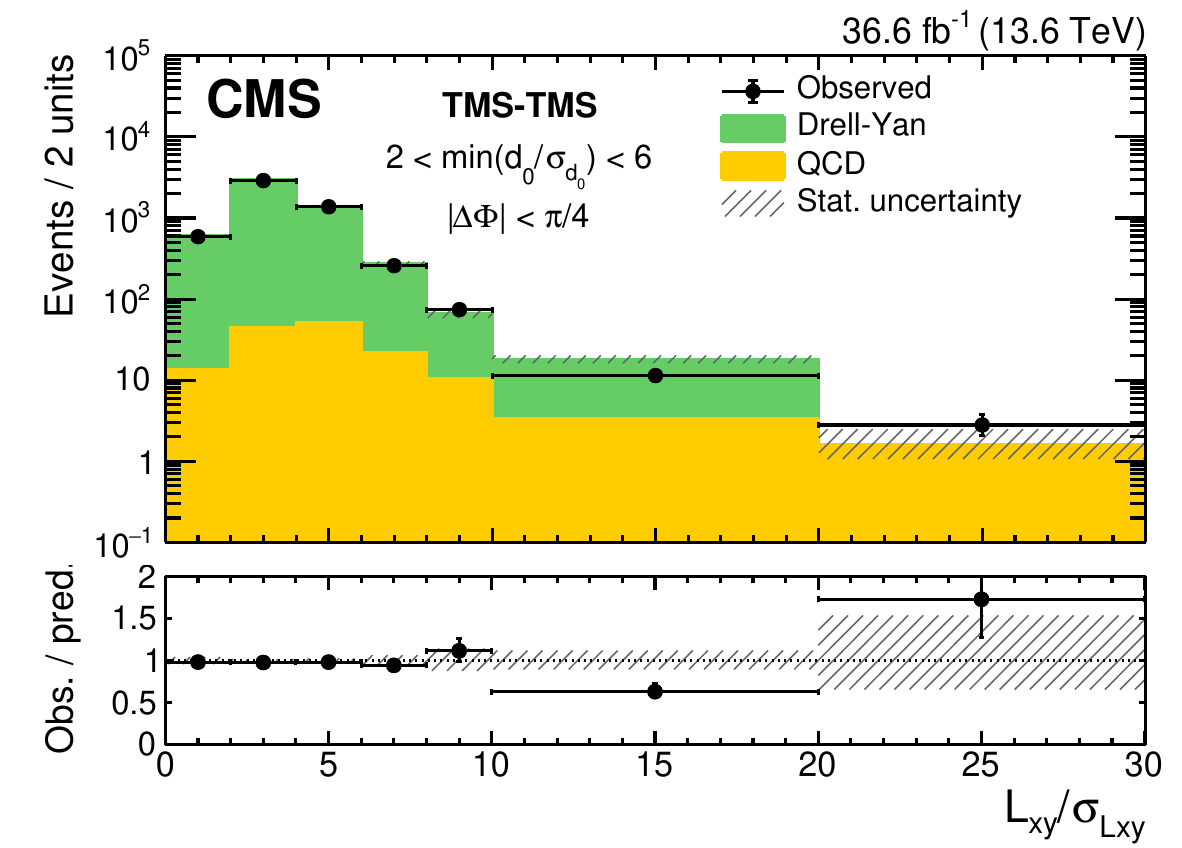}
  \includegraphics[width=0.95\DSquareWidth]{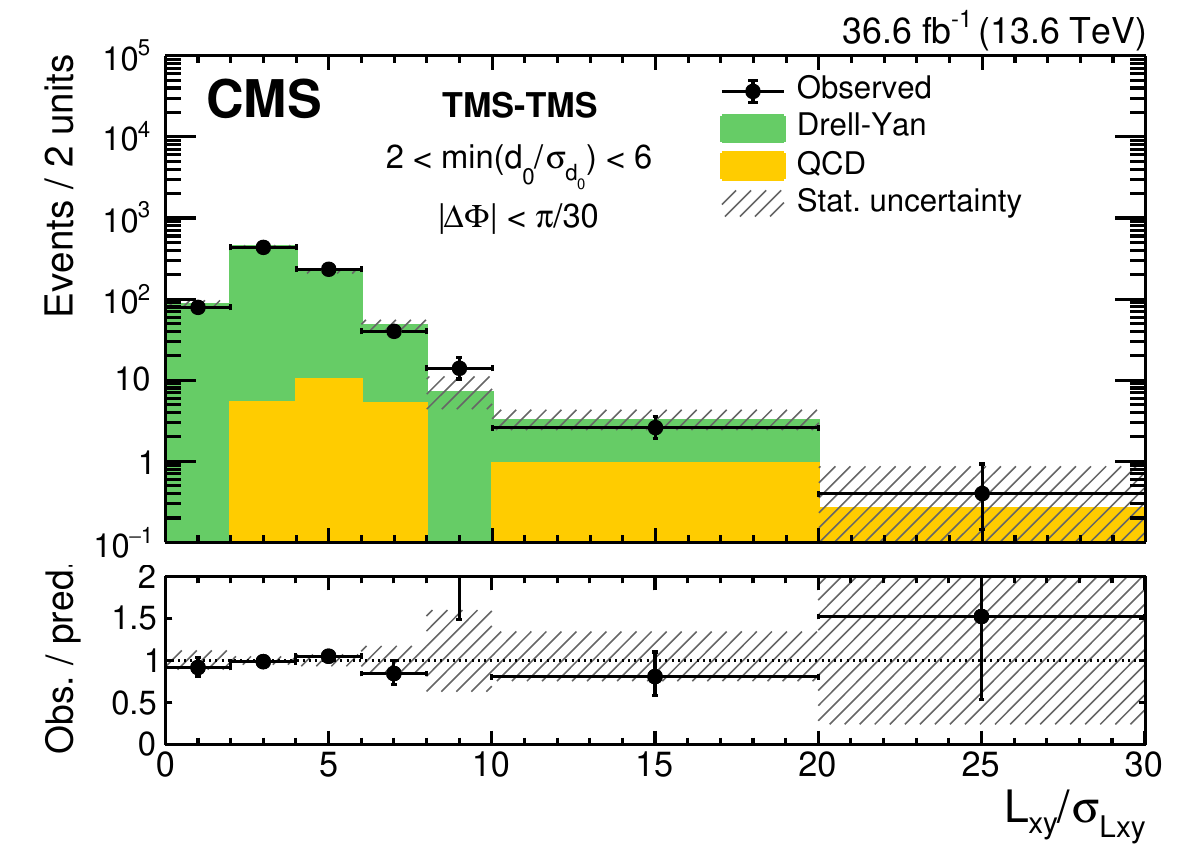}
    \caption{Example of background prediction checks in the TMS-TMS category:
    $\LxySig$ distributions for events with (left) 
    $\DeltaPhiAbs < \pi/4 $ and (right) $\DeltaPhiAbs < \pi/30$
    in the $2 < \text{min}(\dzeroSig) < 6$ validation regions
    compared to the background predictions.  
    The number of observed events (black circles) is overlaid with stacked
    histograms showing the expected numbers of QCD (yellow) and
    DY (green) background events.  
    The last
    bin includes events in the histogram overflow.  The lower panels show the ratio of
    the observed to predicted number of events.  Hatched histograms
    show the statistical uncertainty in the background prediction.}
  \label{fig:Lxysig_DPHI1_PATPAT}
\end{figure}

In the TMS-TMS category, we apply the background evaluation procedure to the
TMS-TMS dimuons in the VR defined by $2 < \text{min}(\dzeroSig) < 6$.  The
comparison of the predicted background and data in bins of $\LxySig$
for two $\DeltaPhiAbsC$ requirements is shown in Fig.~\ref{fig:Lxysig_DPHI1_PATPAT}.  
The expected and observed numbers of events are in agreement in the entire probed
$\LxySig$ range.

\begin{figure}
  \centering
   \includegraphics[width=0.95\DSquareWidth]{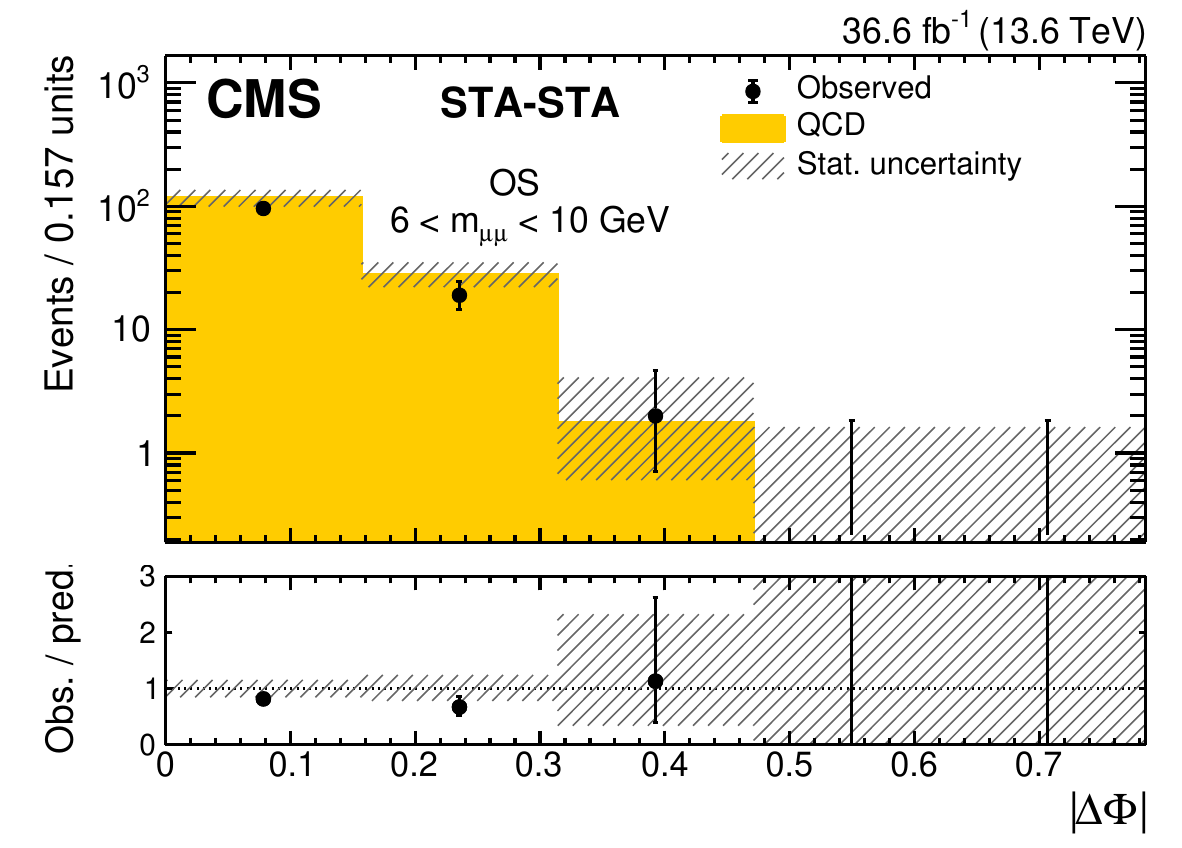}
   \includegraphics[width=0.95\DSquareWidth]{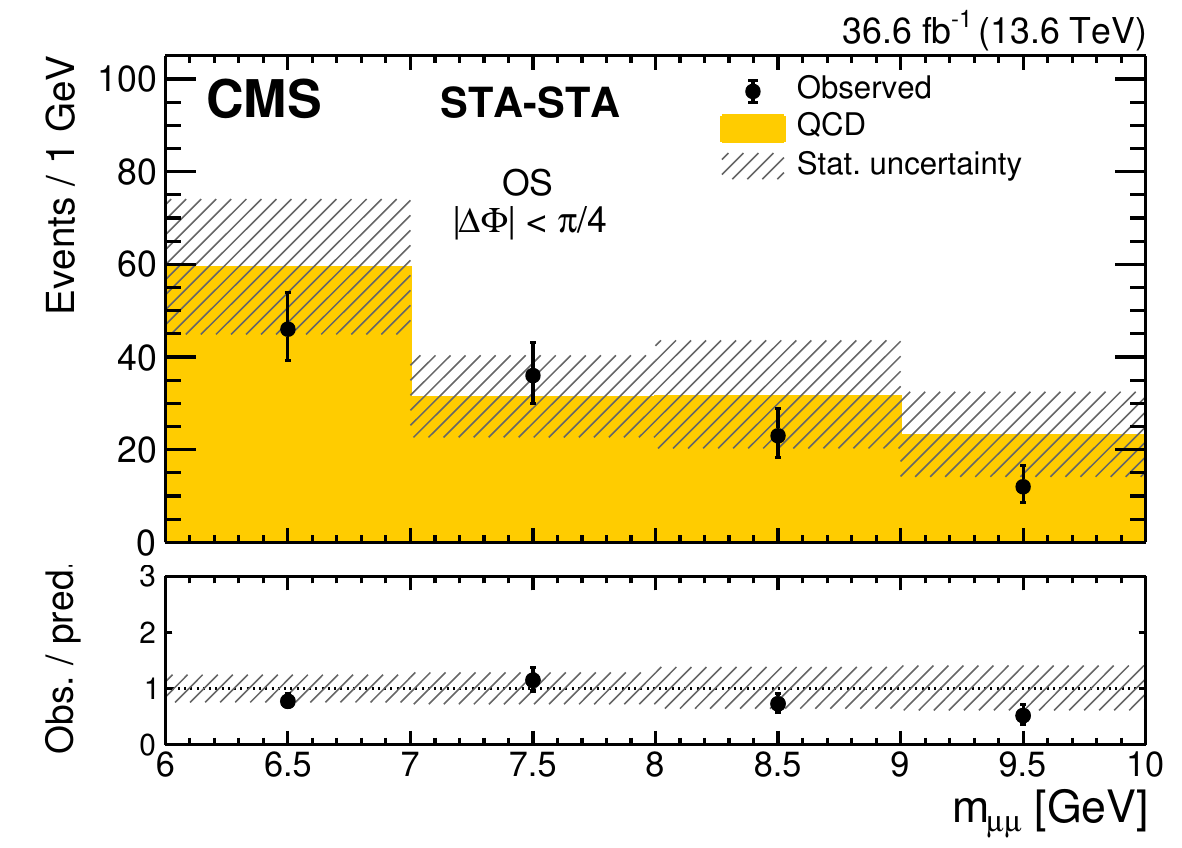}
    \caption{Example of background prediction checks in the STA-STA category: 
    distributions of (left) $\DeltaPhiAbs$ and (right) $\mMuMu$ for dimuons in the low-mass 
     ($6 < \mMuMu < 10 \GeV$) validation region in data (black circles)
     compared to the background predictions (histograms).
    The lower panels show the ratio of the observed to predicted number
    of events.  Hatched histograms show the statistical uncertainty in
    the background prediction.}
   \label{fig:QCD_closure}
 \end{figure}

The evaluation of the $\DeltaPhiAbs$-asymmetric component of the
background, which is particularly important in the STA-STA
category, is tested in the low-mass ($6 < \mMuMu <
10\GeV$) VR where the QCD backgrounds dominate.  Figure~\ref{fig:QCD_closure} shows a comparison
of predicted and measured background in this VR as a function of
$\DeltaPhiAbs$ and $\mMuMu$ for STA-STA dimuons.
The yields in data are found to be consistent with the background
predictions in all $\DeltaPhiAbs$ and \mMuMu intervals.

\begin{figure}
  \centering
  \includegraphics[width=0.95\DSquareWidth]{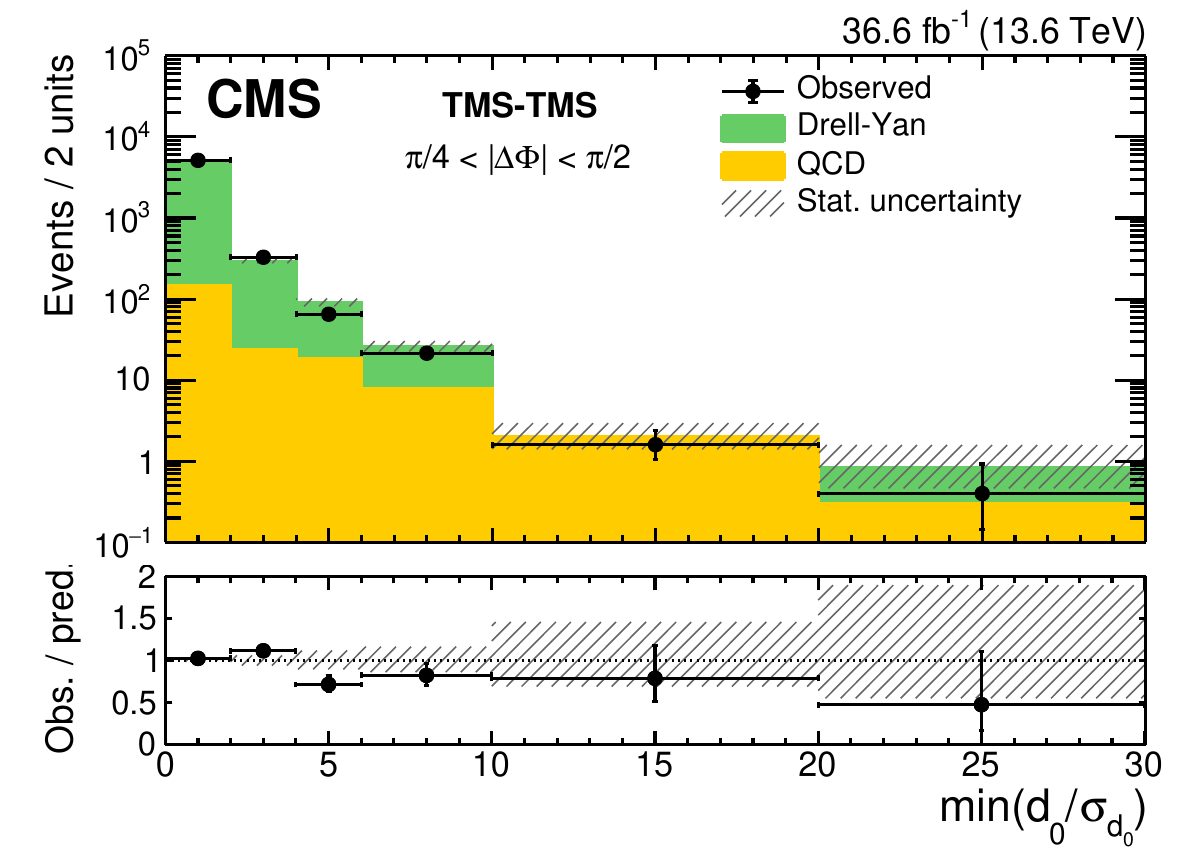}
  \includegraphics[width=0.95\DSquareWidth]{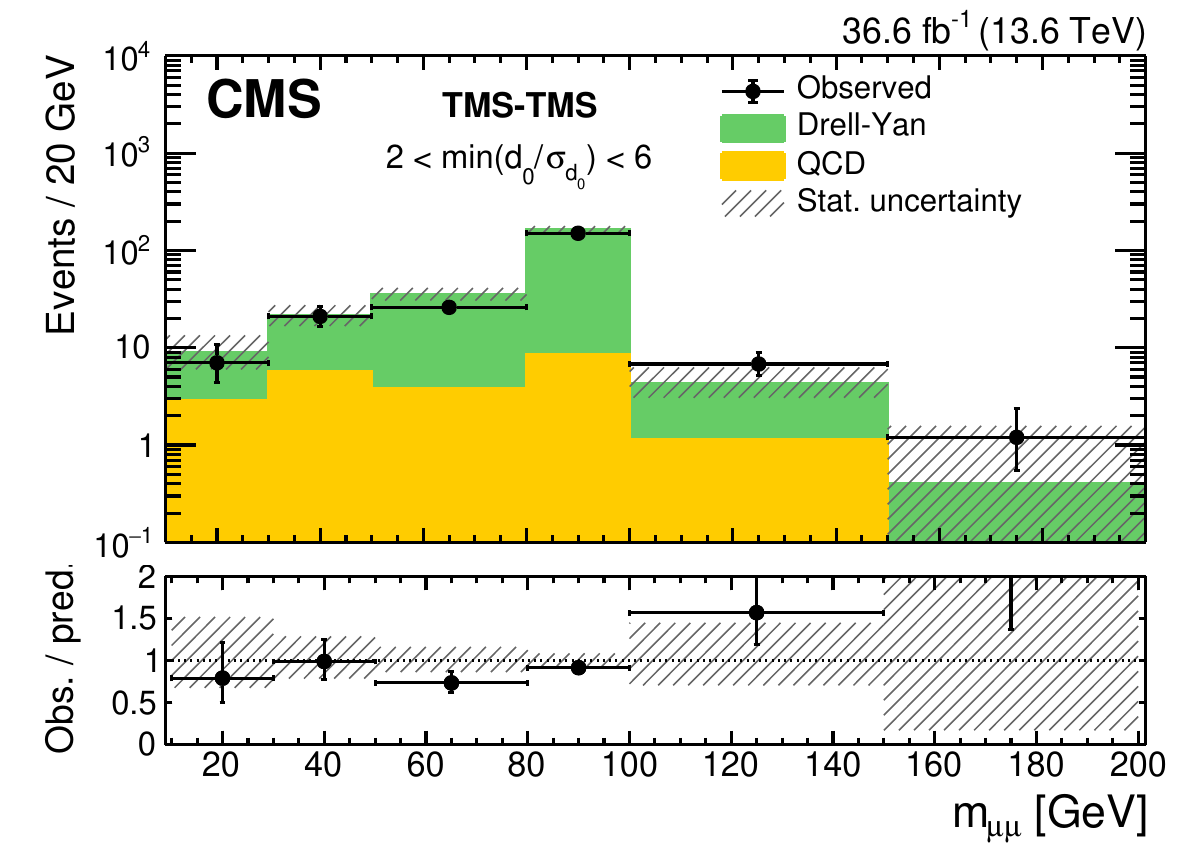}
  \hspace*{-1.3em}
  \caption{Example of background prediction checks in the TMS-TMS category:
    (left) distribution of min($\dzeroSig$) for events in the $\pi/4 < \DeltaPhiAbs < \pi/2$ validation region;
    (right) distribution of \mMuMu for events in the $2 < \text{min}(\dzeroSig) < 6$ validation region.
    The number of observed events (black circles) is overlaid with stacked
    histograms showing the expected numbers of QCD (yellow) and
    DY (green) background events.
    The last bin
    includes events in the histogram overflow.  The lower panels show the ratios
    of the observed to predicted number of events.  Hatched histograms
    show the statistical uncertainty in the background prediction.}
  \label{fig:d0sigpv_DPHI2_closure}
\end{figure}

Finally, to ensure the validity of the method at different
values of the main discriminating variables in the TMS-TMS
category, the validation checks are performed in bins of min($\dzeroSig$)
and $\mMuMu$ of TMS-TMS dimuons.  Such checks include comparisons in the $\dzeroSig$
sideband ($2 <\dzeroSig < 6$) in the $\DeltaPhiAbs<\pi/4$ SR, as well
as those in the entire $\dzeroSig$ range in the $\DeltaPhiAbs$ sideband,
$\pi/4 < \DeltaPhiAbs < \pi/2$.  In the latter, the region with
$\pi/4 < \DeltaPhiAbs < \pi/2$ is used as a signal-free proxy for the
$\DeltaPhiAbs < \pi/4$ SR.  The background evaluation
procedure is applied to the OS and SS dimuons in the
$\DeltaPhiAbs$-symmetric region, $\pi/2 < \DeltaPhiAbs < 3\pi/4$, as
well as SS dimuons with $\pi/4 < \DeltaPhiAbs < \pi/2$.  The
comparisons of the predicted background and data as a function of
min($\dzeroSig$) and \mMuMu for TMS-TMS dimuons in these VRs are shown in
Fig.~\ref{fig:d0sigpv_DPHI2_closure}.  The observed and expected
numbers of events are consistent within statistical uncertainties.

\section{Systematic uncertainties affecting signal} \label{sec:systunc_signal}

The modeling of signal efficiencies in simulation was extensively
studied in the Run 2 search~\cite{EXO-21-006}.  Since no changes
affecting this analysis were made in the CMS detector or in the muon
reconstruction algorithms between Run 2 and Run 3, many of the
systematic uncertainties and data-to-simulation corrections are taken
directly from Ref.~\cite{EXO-21-006}.  New studies include modeling of
signal efficiencies related to the new trigger algorithms and modified
event selection criteria, such as the STA muon isolation.  The studies
are performed for each  dimuon category separately, using dedicated data samples.
Unless stated otherwise, we consider sources of uncertainties to be
uncorrelated between different categories.

In both categories, the dominant systematic uncertainties and
the largest data-to-simu\-lation corrections arise from muon
identification, muon reconstruction, and trigger efficiencies.  At
small displacements, the systematic effects related to these
efficiencies are examined as a function of muon \pt and $\eta$ by
applying the ``tag-and-probe'' method~\cite{Chatrchyan:2012xi} to
prompt muons from $\JPsi$ meson and $\PZ$ boson decays.
The evolution of
efficiencies as a function of displacement is studied 
using cosmic ray muons
and muons from decays of nonprompt $\JPsi$ mesons. 
These studies yield corrections to the simulated signal yields that range, depending on the signal sample, 
from 0.75 to unity for the STA-STA category
and from 0.83 to unity for the TMS-TMS category.
The overall uncertainty increases continuously with displacement, 
from ${\approx}5\%$ for prompt-like muons to ${\approx}15\%$ for muons with $\dzero=100\cm$, 
reflecting the precision of studies using cosmic ray muons. 
Since a large part of this uncertainty
stems from the evaluation of efficiencies of trigger algorithms common to both dimuon categories, it is taken as correlated between
the categories.

Dedicated studies were performed to examine the effect of the $\dzero$
thresholds and looser muon \pt requirements of the new displaced
dimuon triggers.  The efficiency of the Run 3 (2022, L3) triggers is
evaluated using dimuons from decays of nonprompt $\JPsi$
mesons selected by the triggers that use jets and event $\pt$ imbalance.
As a result of these studies, the simulated yields of signal
events selected 
by this set of triggers are scaled by a factor of
0.95, and an uncertainty of 5\% is assigned. The efficiency of the Run
3 (2022, L2) triggers is studied using cosmic ray muons.  While no
systematic biases are observed, the largest data-simulation
difference of 3\% in bins of $\dzero$ is assigned as a systematic
uncertainty in the efficiency of these triggers.  These corrections and
uncertainties are included in the overall corrections and
uncertainties discussed in the previous paragraph.

The accuracy of the modeling of the muon isolation requirements is
assessed using muons from $\PZ$ boson decays and, in the case of STA
muons, cosmic ray muons selected in $\Pp\Pp$ collision events.
Based on the results of these studies,
a systematic uncertainty of 6 (2)\% is assigned to the efficiency of STA (TMS) muons.
The uncertainty in the Higgs boson production cross section at 13.6\TeV
amounts to $^{+5\%}_{-7\%}$~\cite{LHCHWG136TeV}.
The remaining systematic uncertainties, \eg, those related to
muon \pt resolution and dimuon vertex reconstruction, are less than 5\%. 
The uncertainty in the integrated luminosity is 2.3\%~\cite{CMS:2023qyl}.
The uncertainty in the signal efficiency due to pileup modeling is 2\%.  Both luminosity and pileup uncertainties are correlated
among the dimuon categories.

\section{Results} \label{sec:results}

\begin{figure}
  \centering
  \includegraphics[width=0.97\DSquareWidth]{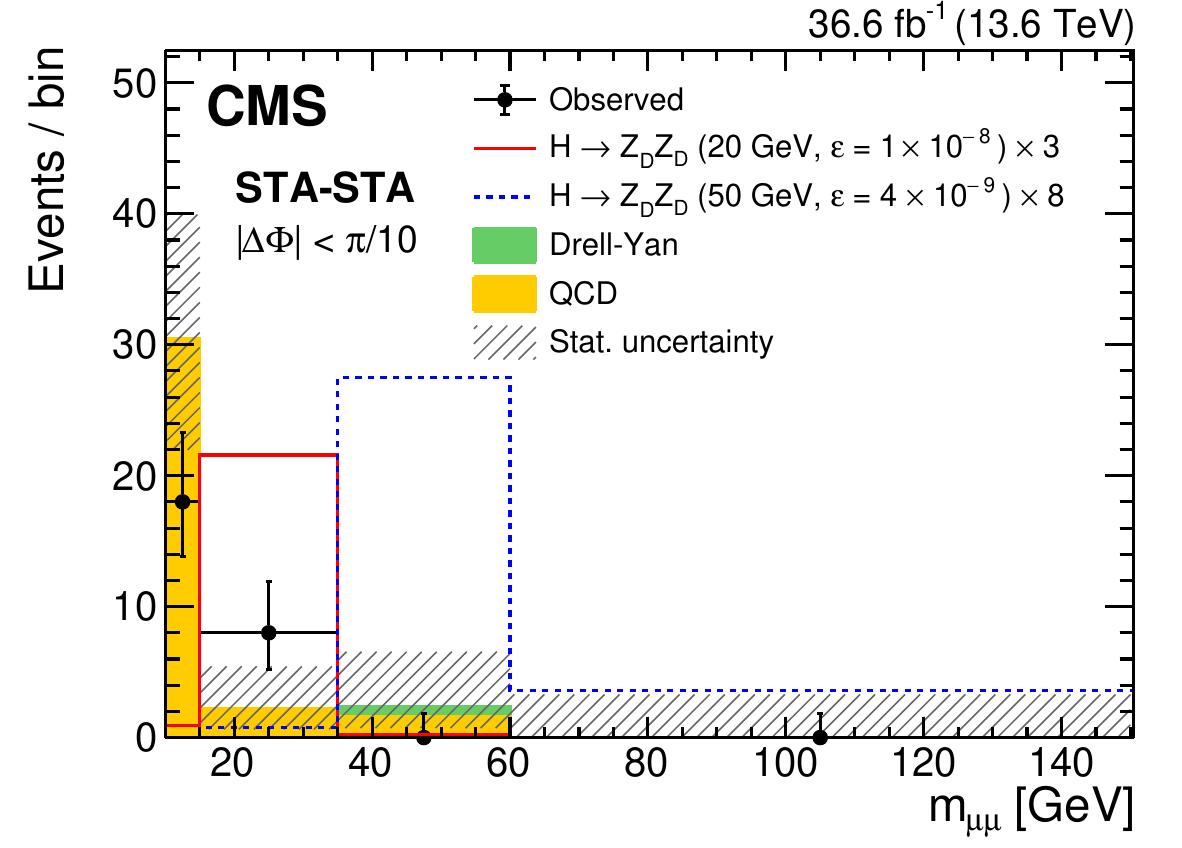}
  \includegraphics[width=0.97\DSquareWidth]{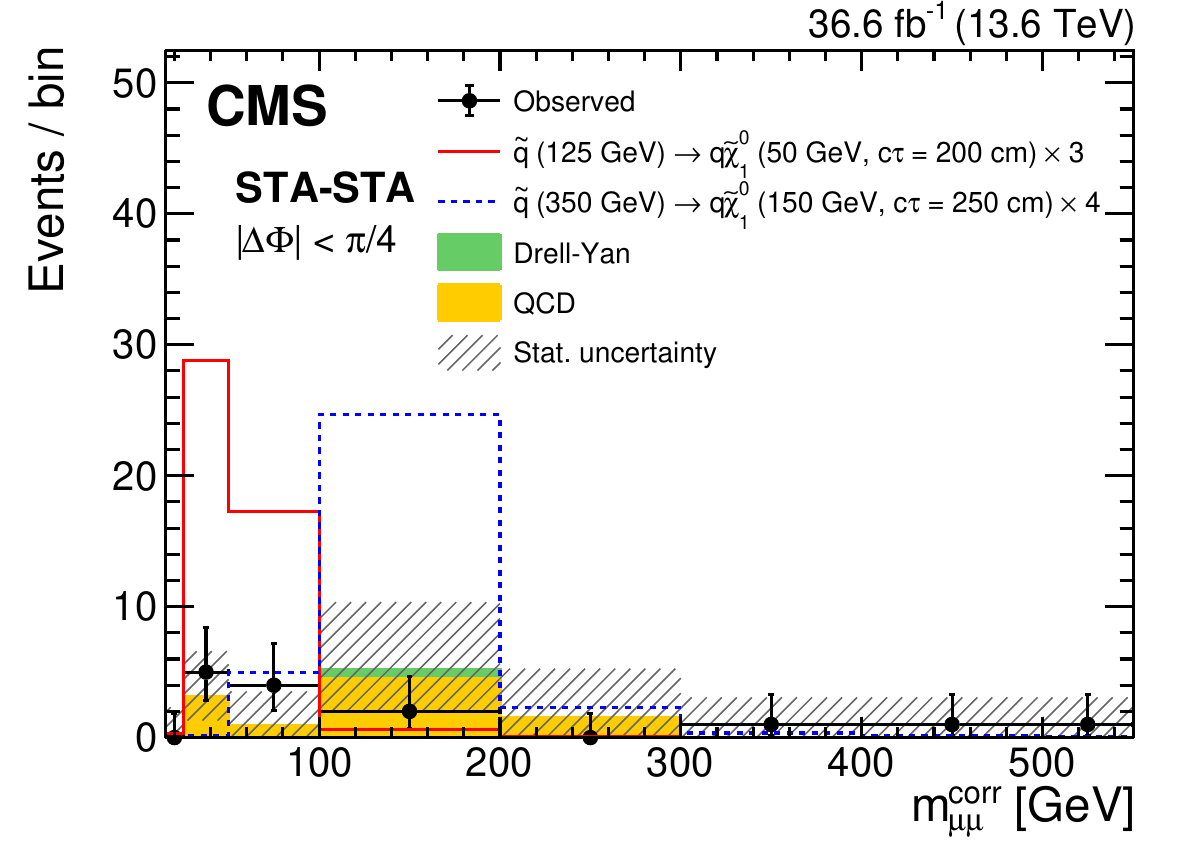}
  \caption{Comparison of the observed (black points) and
    expected (histograms) numbers of events in nonoverlapping (left)
    \mMuMu and (right) \mMuMuCorr intervals in the STA-STA dimuon
    category, in the signal regions optimized for the (left) HAHM and (right) RPV SUSY model.
    Yellow and green stacked filled histograms represent mean expected background contributions from QCD and DY, respectively,
    while statistical uncertainties in the total expected background are shown as hatched histograms.  
    Signal contributions expected from simulated signals indicated in the legends are shown in red and blue. 
    Their yields are set to the corresponding
    median expected 95\% \CL exclusion limits obtained from the ensemble of both dimuon categories, scaled up as indicated in the legend to improve visibility.
    The last bin includes events in the histogram overflow.
    }
    \label{fig:obs_vs_exp_invmass_STASTA}
\end{figure}

The numbers of observed events and the predicted background yields in
the STA-STA category are shown in Fig.~\ref{fig:obs_vs_exp_invmass_STASTA}
in representative \mMuMu and
\mMuMuCorr intervals.  The width of the mass intervals and the
$\DeltaPhiAbs$ and \mMuMu requirements for events in
Fig.~\ref{fig:obs_vs_exp_invmass_STASTA} (left) and
Fig.~\ref{fig:obs_vs_exp_invmass_STASTA} (right) correspond to those
chosen, respectively, for the study of the HAHM and of the RPV SUSY
model.
For illustrative purposes, signal distributions at the level of the median expected exclusion
limits at 95\% confidence level (\CL) obtained from the ensemble of both dimuon categories in the background-only scenario are also shown.
The numbers of observed events are consistent with background predictions.
As expected for background events, most of the observed events have low \mMuMu.

\begin{figure}
  \centering
  \includegraphics[width=0.95\DSquareWidth]{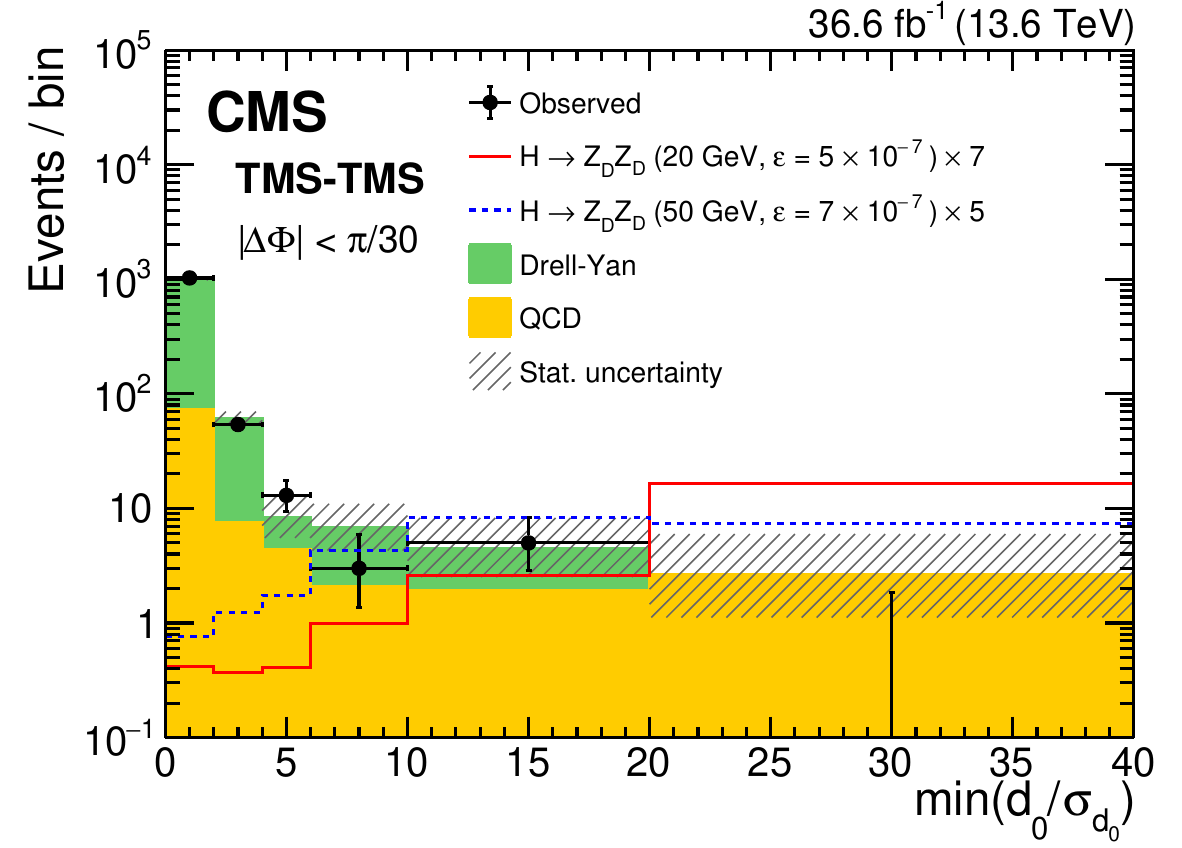}
  \includegraphics[width=0.95\DSquareWidth]{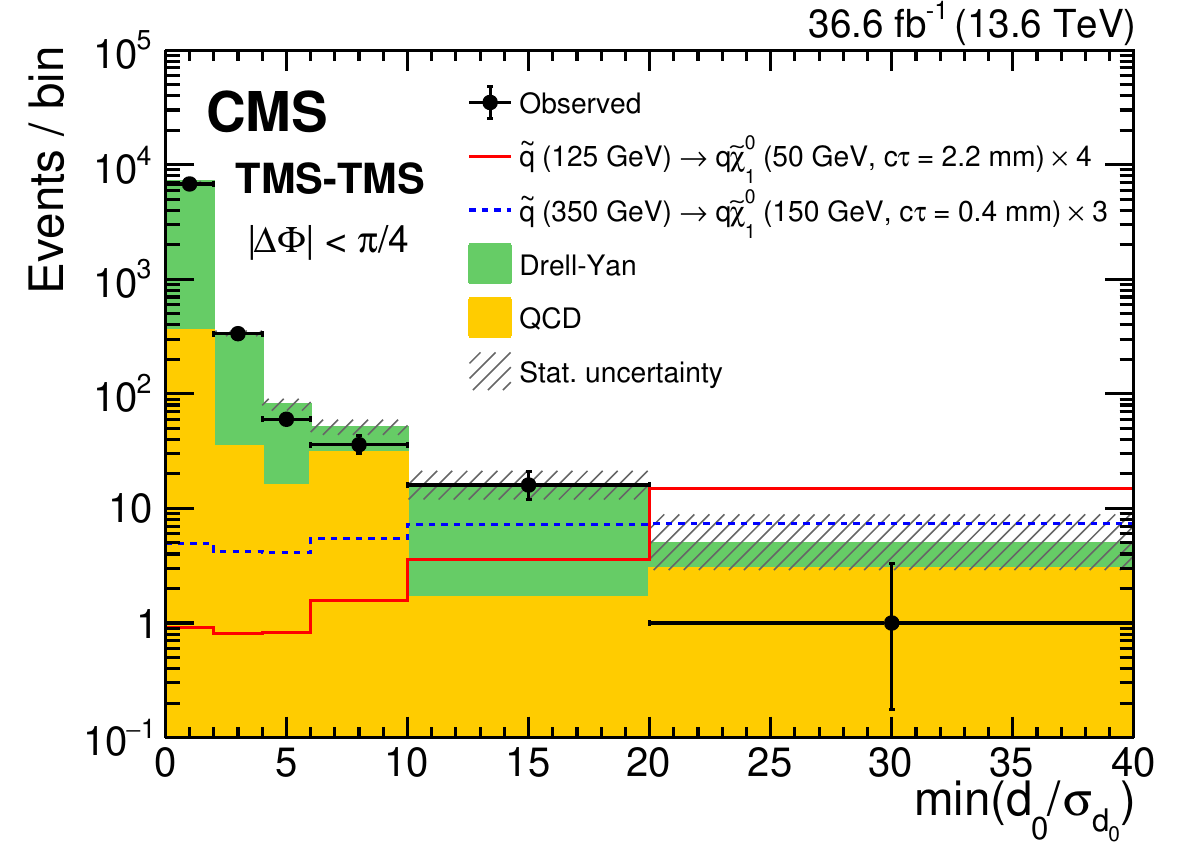}
  \caption{Distributions of $\text{min}(\dzeroSig)$ for TMS-TMS
    dimuons with (left) $\DeltaPhiAbs < \pi/30$ and (right)
    $\DeltaPhiAbs < \pi/4$, for events in all mass intervals combined,
    for both the validation ($\min(\dzeroSig) < 6$) and signal
    ($\min(\dzeroSig) > 6$) regions.
    The number of observed events (black circles) is overlaid with the
    stacked histograms showing the expected numbers of QCD (yellow)
    and DY (green) background events.
    Statistical uncertainties in the total expected background are
    shown as hatched histograms.  Signal contributions expected from
    simulated signals indicated in the legends are shown in red and
    blue.  Their yields are set to the corresponding median expected
    95\% \CL exclusion limits obtained from the ensemble of both
    dimuon categories, scaled up as indicated in the legend to improve
    visibility.  Events are required to satisfy all nominal selection
    criteria with the exception of the $\dzeroSig$ requirement.  The
    last bin includes events in the histogram overflow.}
  \label{fig:d0sig_IPHI1_PATPAT}
\end{figure}

\begin{figure}
 \centering
 \includegraphics[width=2.0\DSquareWidth]{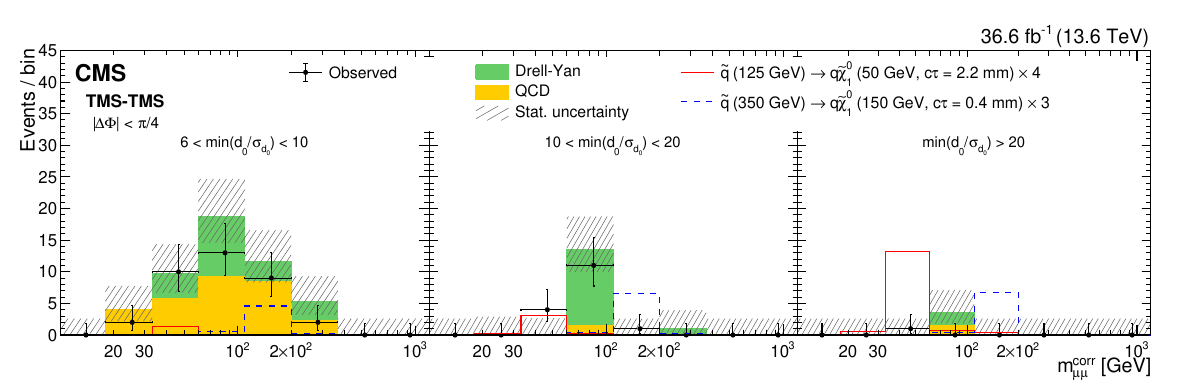}
  \caption{Comparison of observed and expected numbers of events in bins of \mMuMuCorr
    in the TMS-TMS dimuon category, in the signal regions optimized for
    the RPV SUSY model.
    The number of observed events (black circles) is overlaid with the
    stacked filled histograms showing the expected numbers of QCD (yellow)
    and DY (green) background events in bins of
    $\mMuMuCorr$ in three min($\dzeroSig$) bins: (left) 6--10, (center) 10--20, and (right) ${>}20$. 
    Hatched histograms show statistical uncertainties in the total
    expected background.
    Contributions expected from signal events predicted by the RPV SUSY model
    with the parameters indicated in the legends
    are shown as red and blue histograms.
    Their yields are set to the corresponding
    median expected 95\% \CL exclusion limits obtained from the ensemble of both dimuon categories, scaled up as indicated in the legend to improve visibility.
    The last bin includes events in the histogram overflow.}
  \label{fig:mass_IPHI1_PATPAT_RPV}
\end{figure}

\begin{figure}
 \centering
 \includegraphics[width=2.0\DSquareWidth]{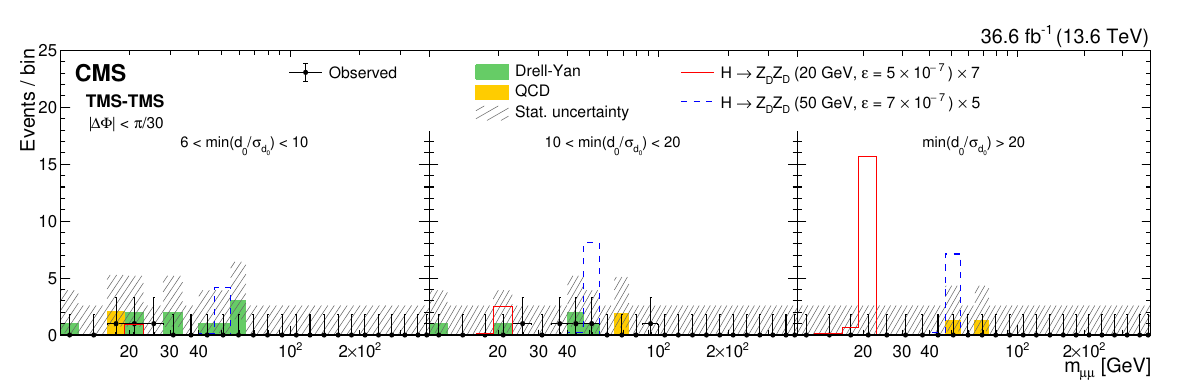}
  \caption{Comparison of observed and expected numbers of events in bins of \mMuMu
    in the TMS-TMS dimuon category, in the signal regions optimized for
    the HAHM.
    The number of observed events (black circles) is overlaid with the
    stacked histograms showing the expected numbers of QCD (yellow)
    and DY (green) background events in bins of
    $\mMuMu$ in three min($\dzeroSig$) bins: (left) 6--10, (center) 10--20, and (right) ${>}20$.  
    Hatched histograms show statistical uncertainties in the total
    expected background.
    Signal contributions expected from simulated $\PSMHiggs \to \PZD \PZD$ events with
    the parameters indicated in the legends
    are shown as red and blue histograms.
    Their yields are set to the corresponding median expected 95\% \CL exclusion limits obtained from the ensemble of both dimuon categories, scaled up as indicated in the legend to improve visibility.
    The last bin includes events in the histogram overflow.}
  \label{fig:mass_IPHI2_PATPAT_HAHM}
\end{figure}

The numbers of observed events and the predicted background yields in
the TMS-TMS category are shown in
Figs.~\ref{fig:d0sig_IPHI1_PATPAT}--\ref{fig:mass_IPHI2_PATPAT_HAHM}.
Figure~\ref{fig:d0sig_IPHI1_PATPAT} shows the
distributions of $\text{min}(\dzeroSig)$  for TMS-TMS 
dimuons with (left) $\DeltaPhiAbs < \pi/30$ and (right) $\DeltaPhiAbs < \pi/4$,
for events in all mass intervals combined.
As expected for background events, the events in data are predominantly at low values of $\text{min}(\dzeroSig)$.
Figure~\ref{fig:mass_IPHI1_PATPAT_RPV} shows the distributions of
$\mMuMuCorr$ in three min($\dzeroSig$) bins, 6--10, 10--20, and ${>}20$, for 
dimuons with $\DeltaPhiAbs < \pi/4$, the looser \DeltaPhiAbs requirement used for the study of the RPV SUSY model.
The numbers of events in the SR chosen for the study of the HAHM model, with the tighter requirement $\DeltaPhiAbs < \pi/30$, are smaller (with no more than one event per $\mMuMu$ bin),
as shown in Fig.~\ref{fig:mass_IPHI2_PATPAT_HAHM}.
The numbers of observed events are consistent with background predictions in both SRs.
The largest $\text{min}(\dzeroSig)$ bin, $\text{min}(\dzeroSig) > 20$, contains the lowest experimental background and,
except for the smallest lifetimes, most of the signal predicted by both models.

These results are used to set upper limits on $\mathcal{B}(\PSMHiggs \to \PZD
\PZD)$ in the HAHM model and on the product of the squark-antisquark
production cross section $\sigma(\Pp\Pp\to\PRPVSquark\PASQ)$
and $\mathcal{B}(\squarktoqchi)$
in the RPV SUSY model.
The limit extraction is based on a modified frequentist
approach~\cite{CLS1, CLS2} and uses the CMS statistical
analysis tool~\textsc{Combine}~\cite{CAT-23-001}.  The method 
yielding background predictions in the SR is implemented
using a multibin likelihood, which is a product of Poisson
distributions corresponding to the SR and CRs.  The systematic uncertainties affecting the signal yield are
incorporated as nuisance parameters using log-normal distributions.
The expected and observed upper limits are evaluated through the use
of simulated pseudo-experiments.
For each signal model, the limits are first computed separately in
each dimuon category.
The individual likelihoods are then combined to obtain the limits in
the ensemble of both categories.  For the interpretation in the
framework of the HAHM model, the results obtained in this analysis are
also combined with the results of the Run 2 analysis~\cite{EXO-21-006}
to achieve even stronger bounds on the model parameters.  The
combination of the two results takes into account the different cross
sections of the Higgs boson production in collisions at 13.0 and
13.6\TeV~\cite{Cepeda:2019klc, LHCHWG136TeV}.

The signal efficiencies used in the statistical interpretations of the results are obtained from simulation and further
corrected by the data-to-simu\-lation scale factors described in
Section~\ref{sec:systunc_signal}. They are computed separately for
each dimuon category, mass interval, LLP lifetime, and signal model.
Given the smallness of the expected
background
and taking into account the
selection efficiencies discussed in Section~\ref{sec:selection}, the
introduction of a separate category for events with two dimuons would
not increase the sensitivity of the analysis significantly even in the
most favorable case for the 4$\PGm$ signal events, namely
$\mathcal{B}(\PLLP\to\MuMu\PX) = 1$.  The gain would be negligible for
smaller $\mathcal{B}(\PZD \to \MuMu)$ values predicted
by the HAHM model.  Therefore, no distinction is made between events
with one and two reconstructed dimuons of the same type.  Events with two TMS-TMS
dimuons are assigned to the min($\dzeroSig$) bin encompassing the
larger of the two min($\dzeroSig$) values.

\begin{figure}
  \centering
  \includegraphics[width=0.95\DSquareWidth]{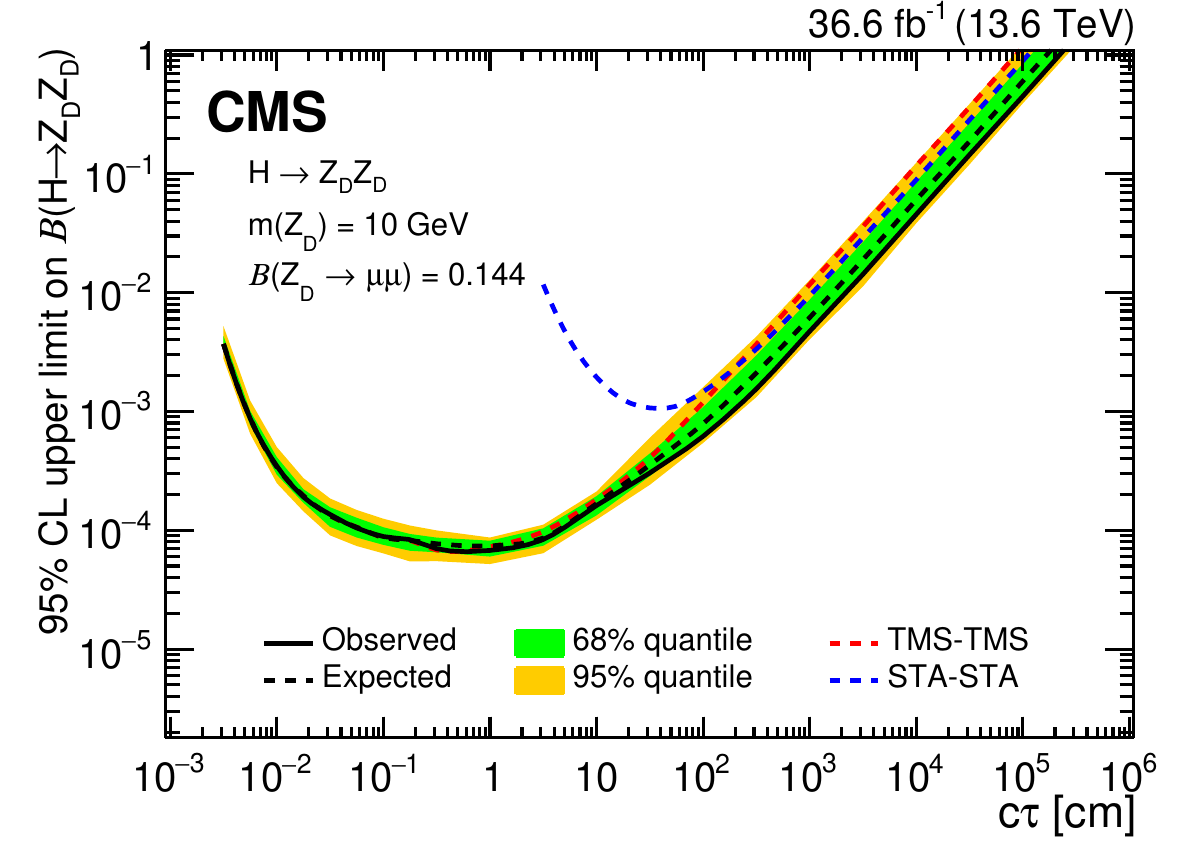}
  \includegraphics[width=0.95\DSquareWidth]{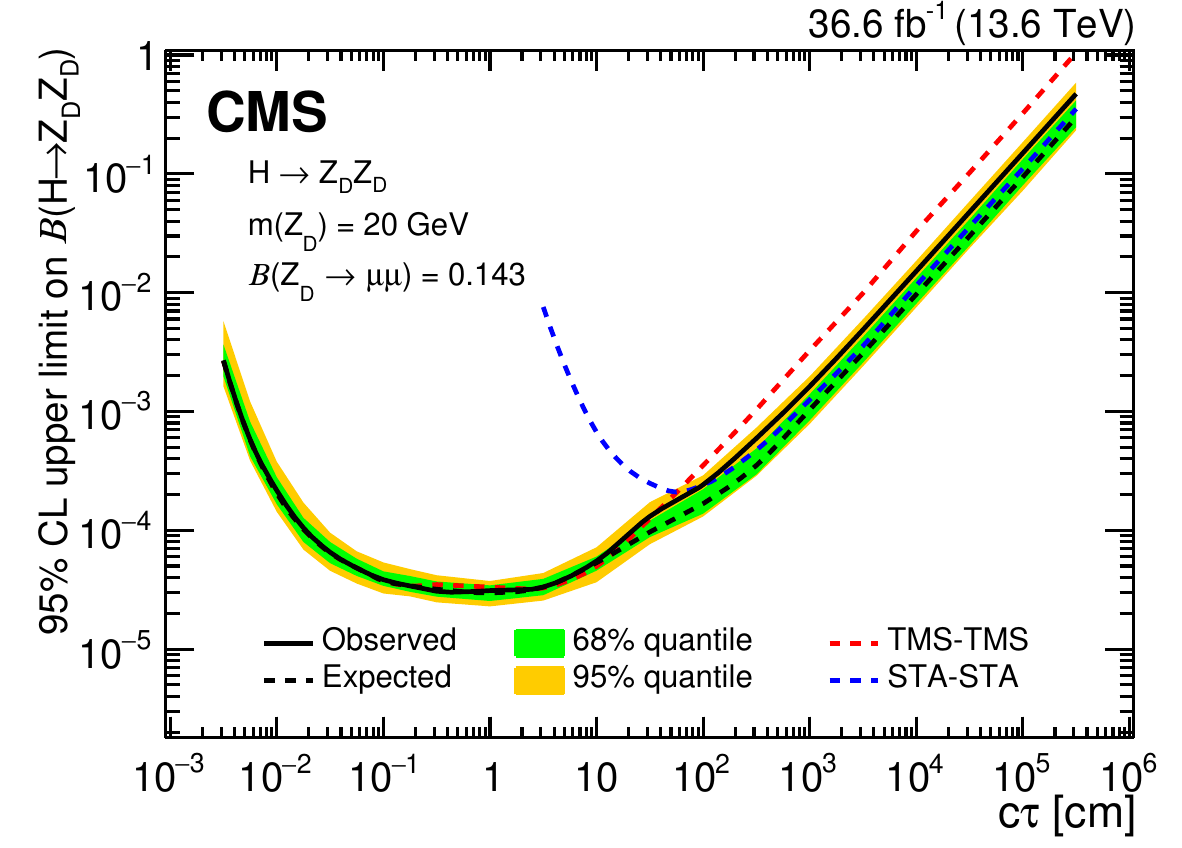}
  \includegraphics[width=0.95\DSquareWidth]{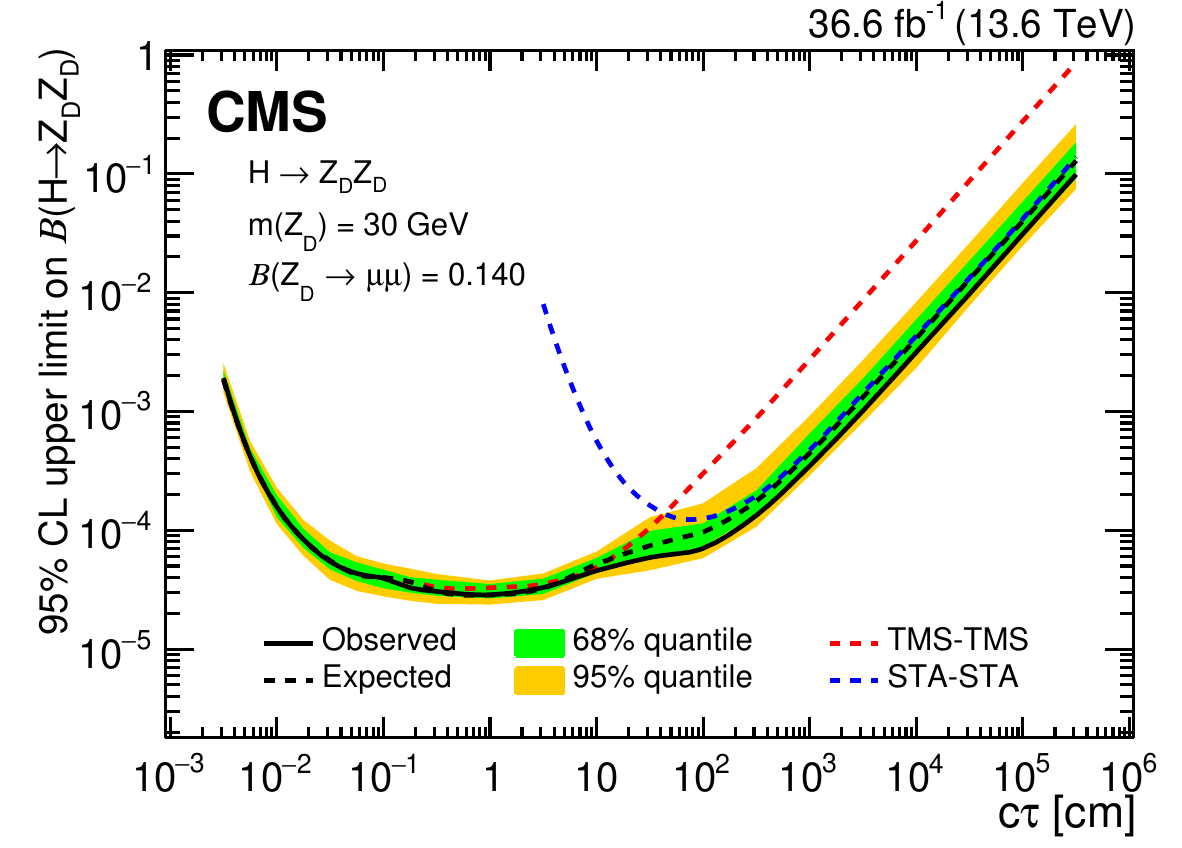}
  \includegraphics[width=0.95\DSquareWidth]{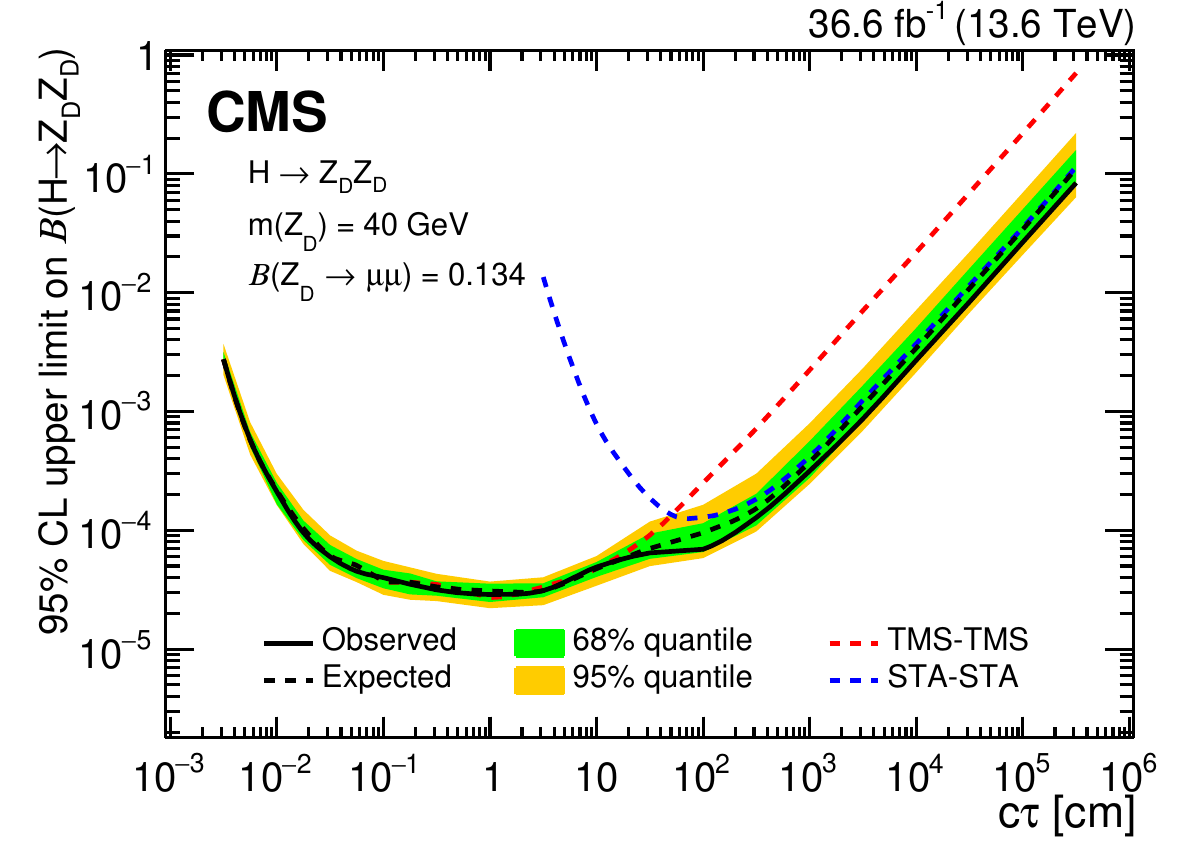}
  \includegraphics[width=0.95\DSquareWidth]{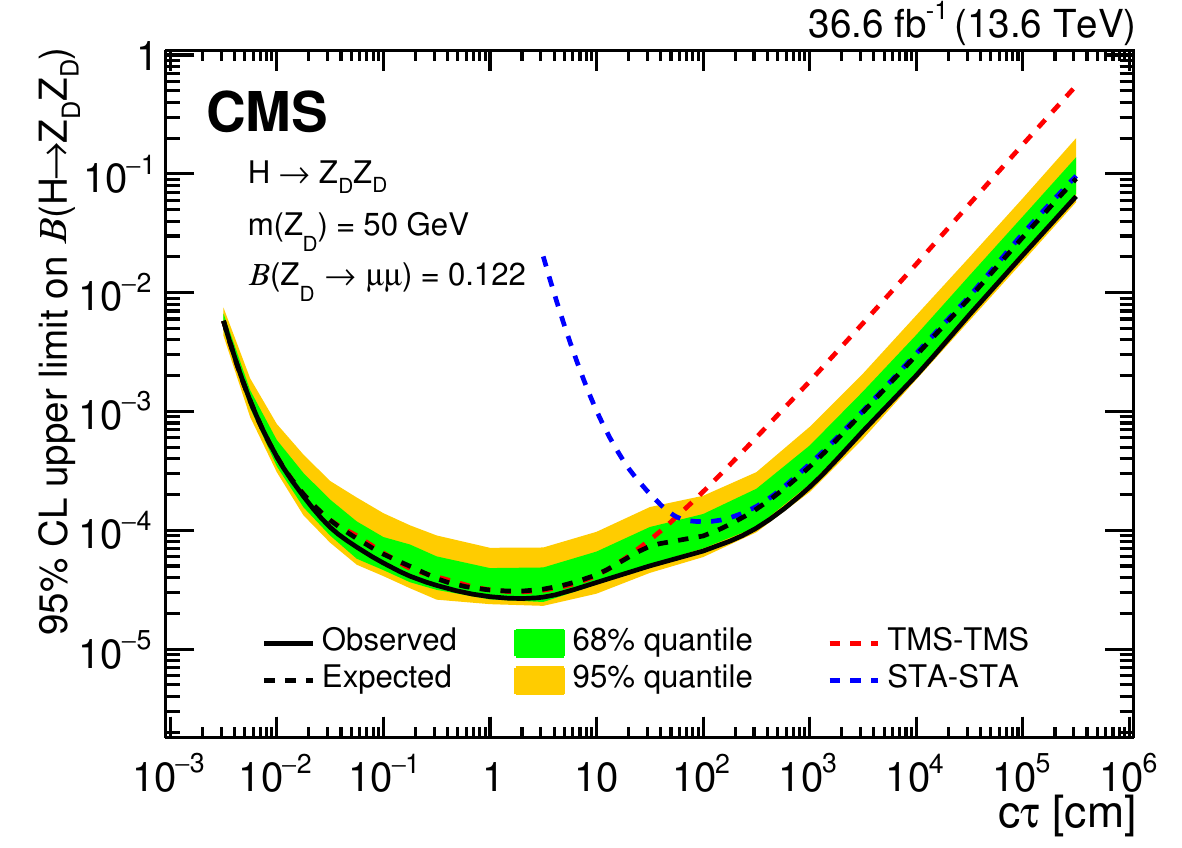}
  \includegraphics[width=0.95\DSquareWidth]{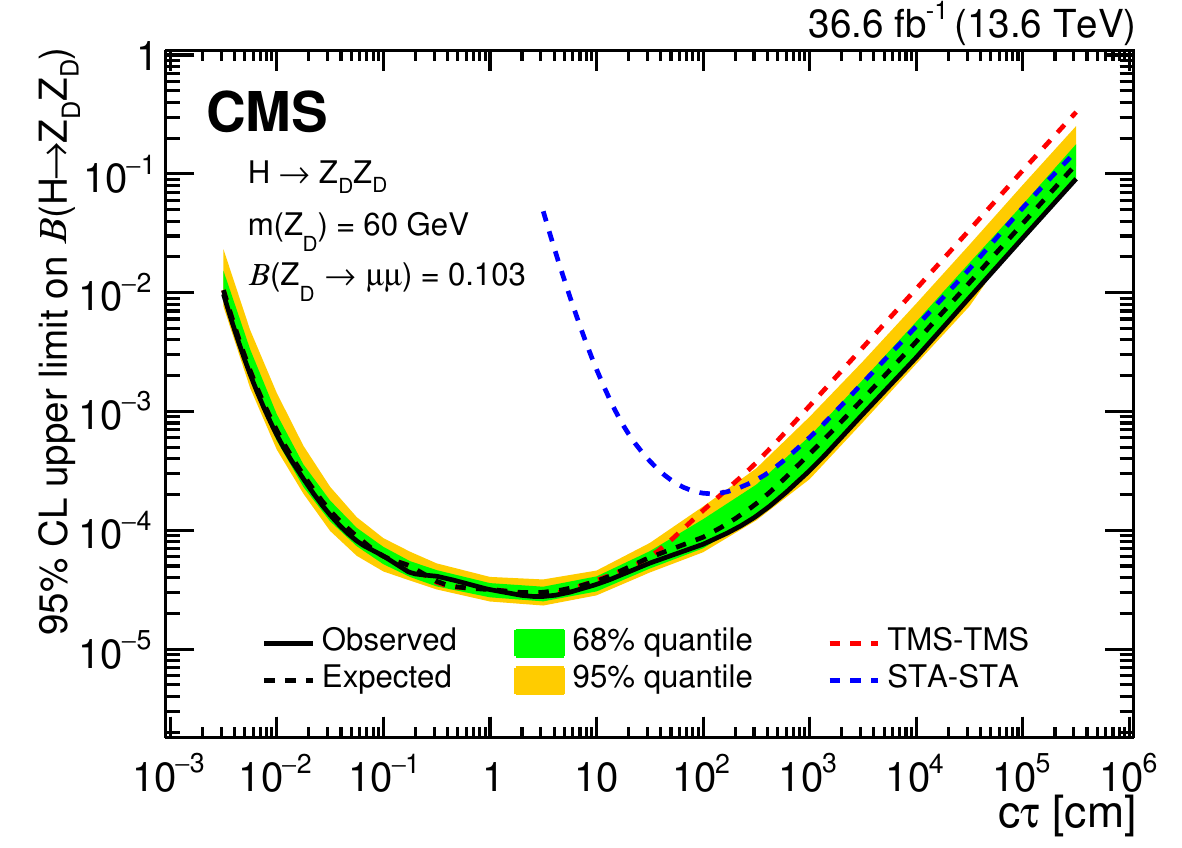}
  \caption{The 95\% \CL upper limits on $\mathcal{B}(\PSMHiggs \to \PZD
    \PZD)$ as a function of $\cTau(\PZD)$ in
    the HAHM model, for $\mZD$ ranging from (upper left) 10\GeV to (lower right)
    60\GeV, in the STA-STA and TMS-TMS dimuon categories in 2022 data
    and their combination.  The median expected limits obtained from
    the STA-STA and TMS-TMS dimuon categories are shown as
    dashed blue and red curves, respectively; the combined
    median expected limits are shown as dashed black curves; and the
    combined observed limits are shown as solid black curves.
    The green and yellow bands correspond, respectively, to the
    68 and 95\% quantiles for the combined expected limits.
    \label{fig:darkphoton_limits}}
\end{figure}

\begin{figure}
  \centering
  \includegraphics[width=0.95\DSquareWidth]{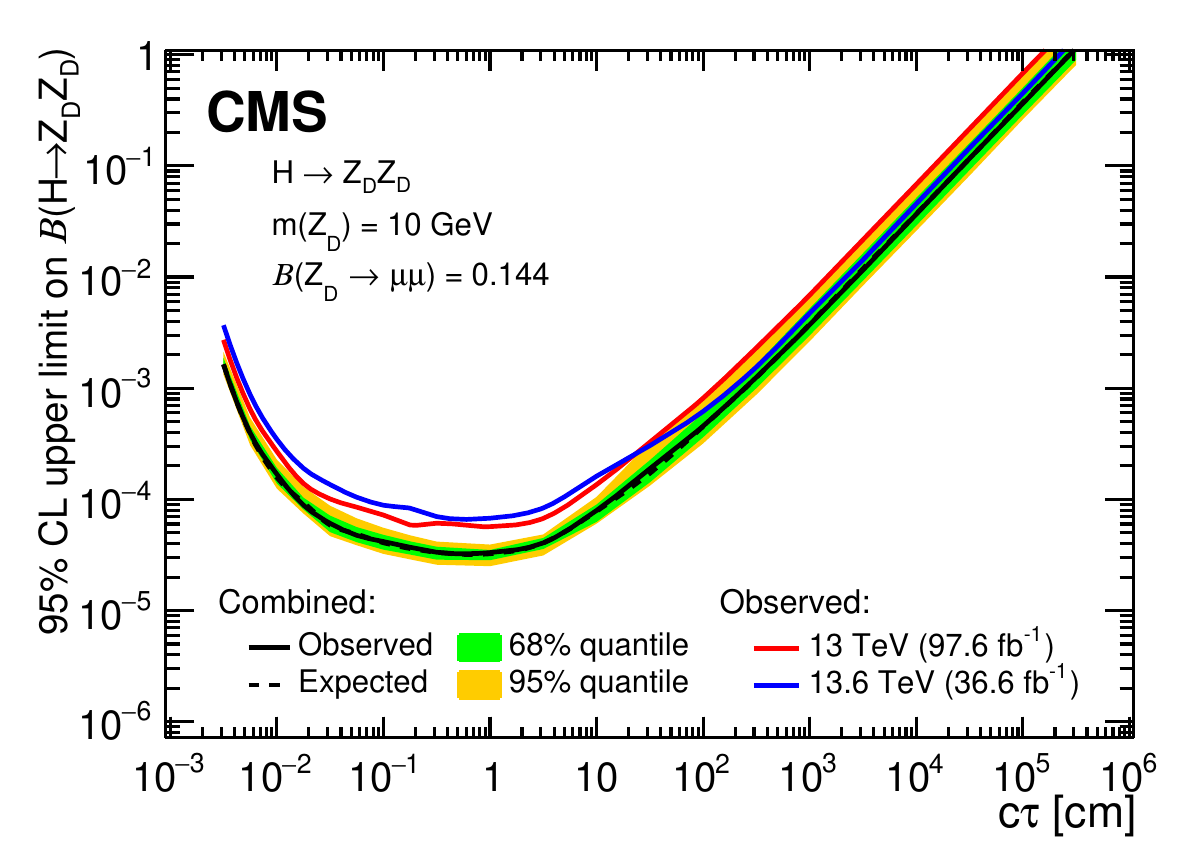}
  \includegraphics[width=0.95\DSquareWidth]{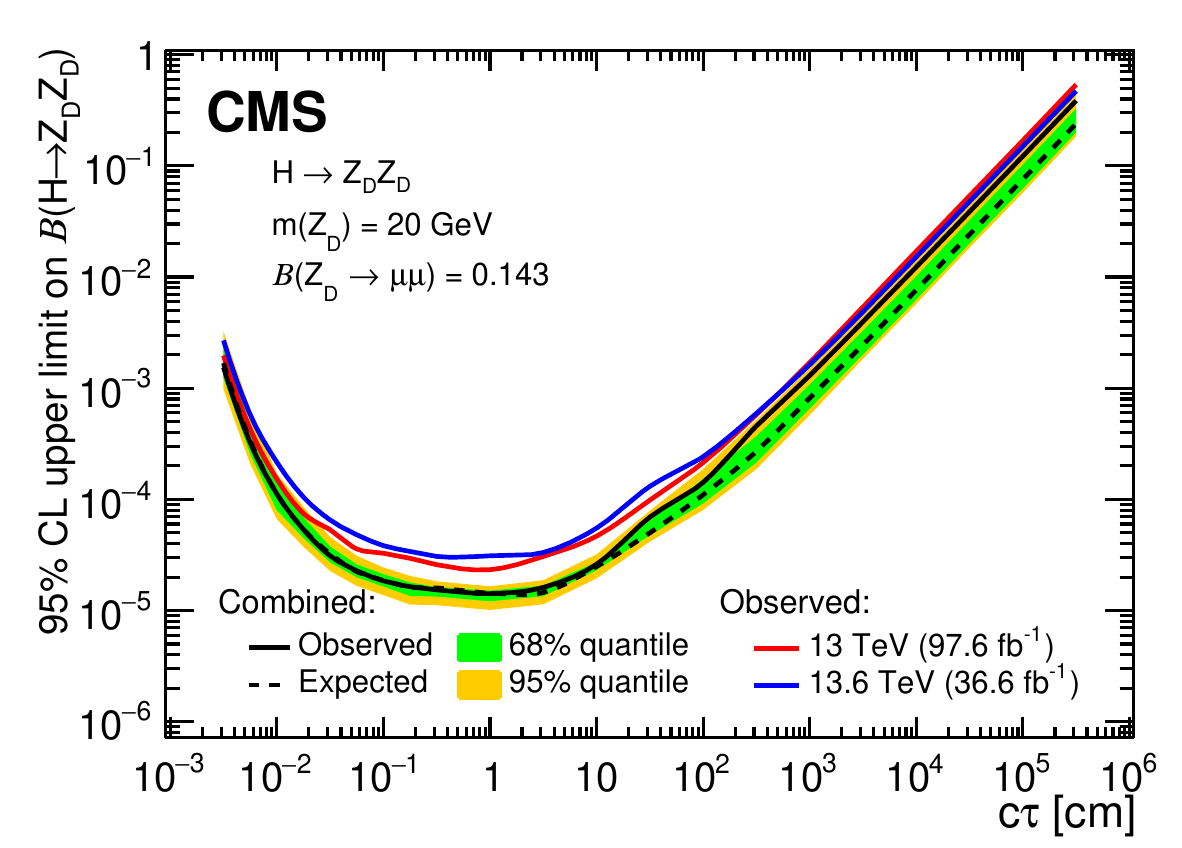}
  \includegraphics[width=0.95\DSquareWidth]{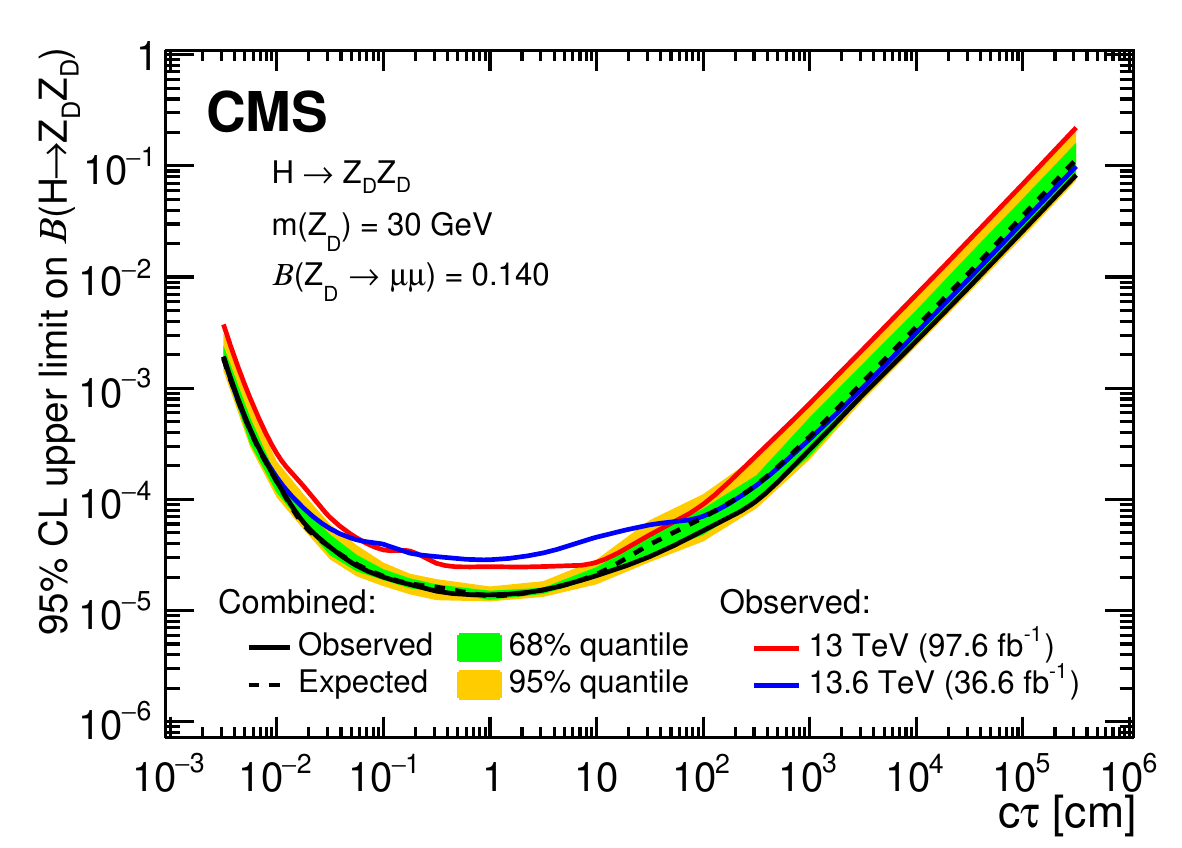}
  \includegraphics[width=0.95\DSquareWidth]{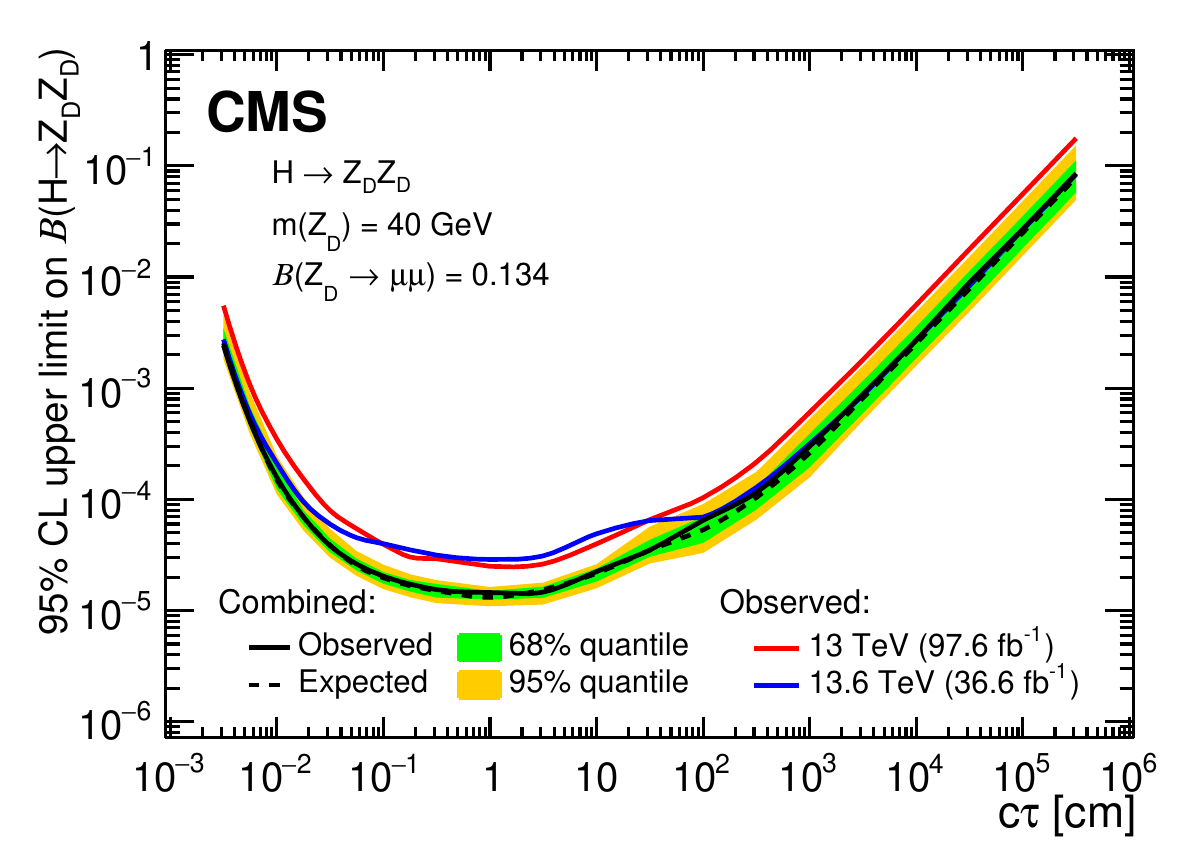}
  \includegraphics[width=0.95\DSquareWidth]{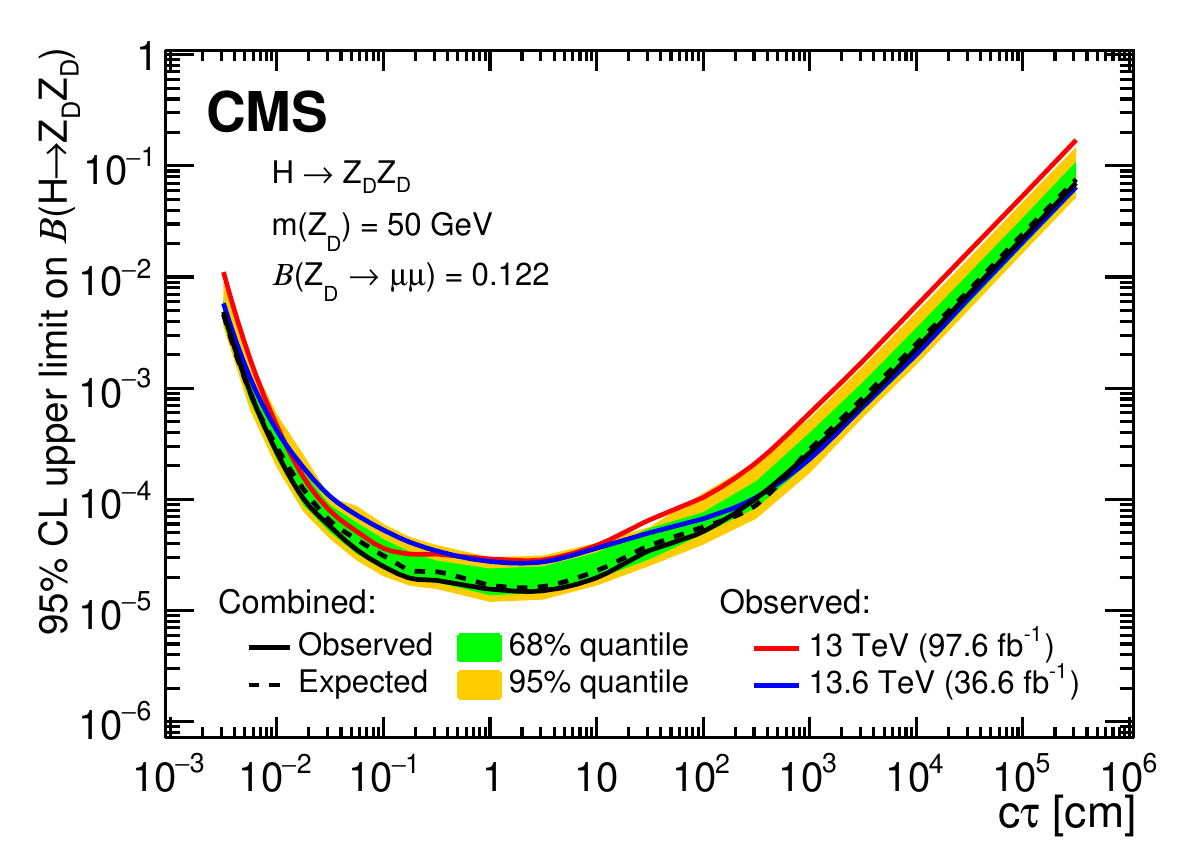}
  \includegraphics[width=0.95\DSquareWidth]{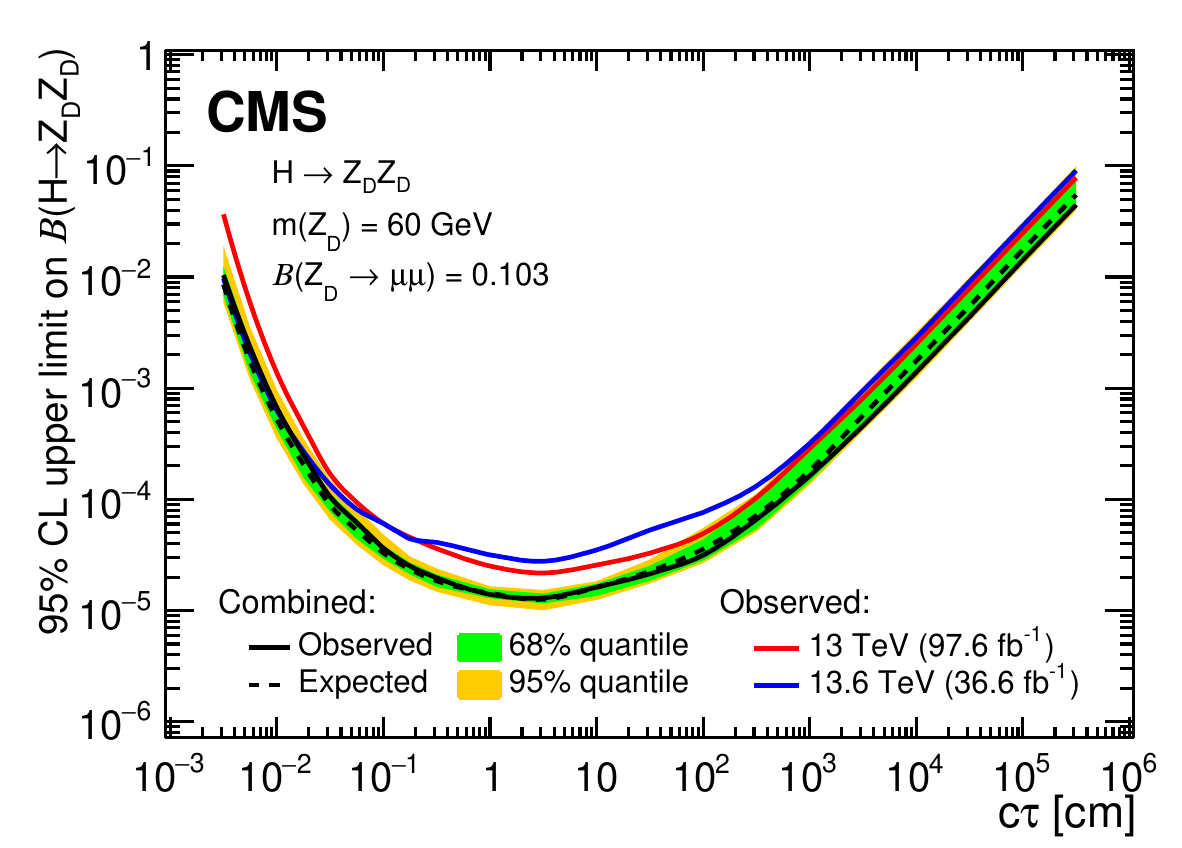}
  \caption{The 95\% \CL upper limits on $\mathcal{B}(\PSMHiggs \to \PZD
    \PZD)$ as a function of $\cTau(\PZD)$ in
    the HAHM model, for $\mZD$ ranging from (upper left) 10\GeV to (lower right)
    60\GeV, obtained in this analysis, the Run~2 analysis~\cite{EXO-21-006}, and their combination.
    The observed limits in this analysis and in the Run 2
    analysis~\cite{EXO-21-006} are shown as blue and red
    curves, respectively; the median combined expected limits are shown as
    dashed black curves; and the combined observed limits are shown as
    solid black curves.  The green and yellow bands correspond,
    respectively, to the 68 and 95\% quantiles for the combined
    expected limits.
    \label{fig:darkphoton_limits_Run23}}
\end{figure}

\begin{figure}
  \centering
  \includegraphics[width=0.95\DSquareWidth]{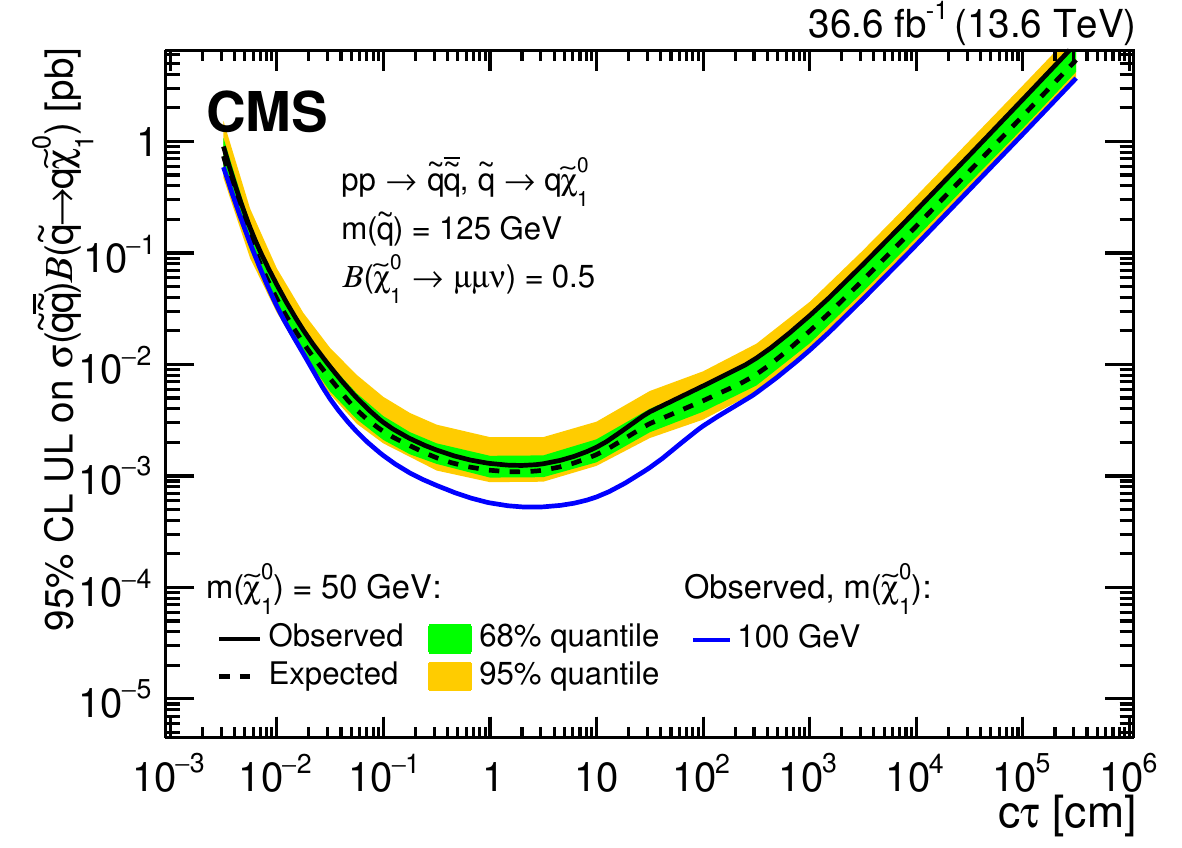}
  \includegraphics[width=0.95\DSquareWidth]{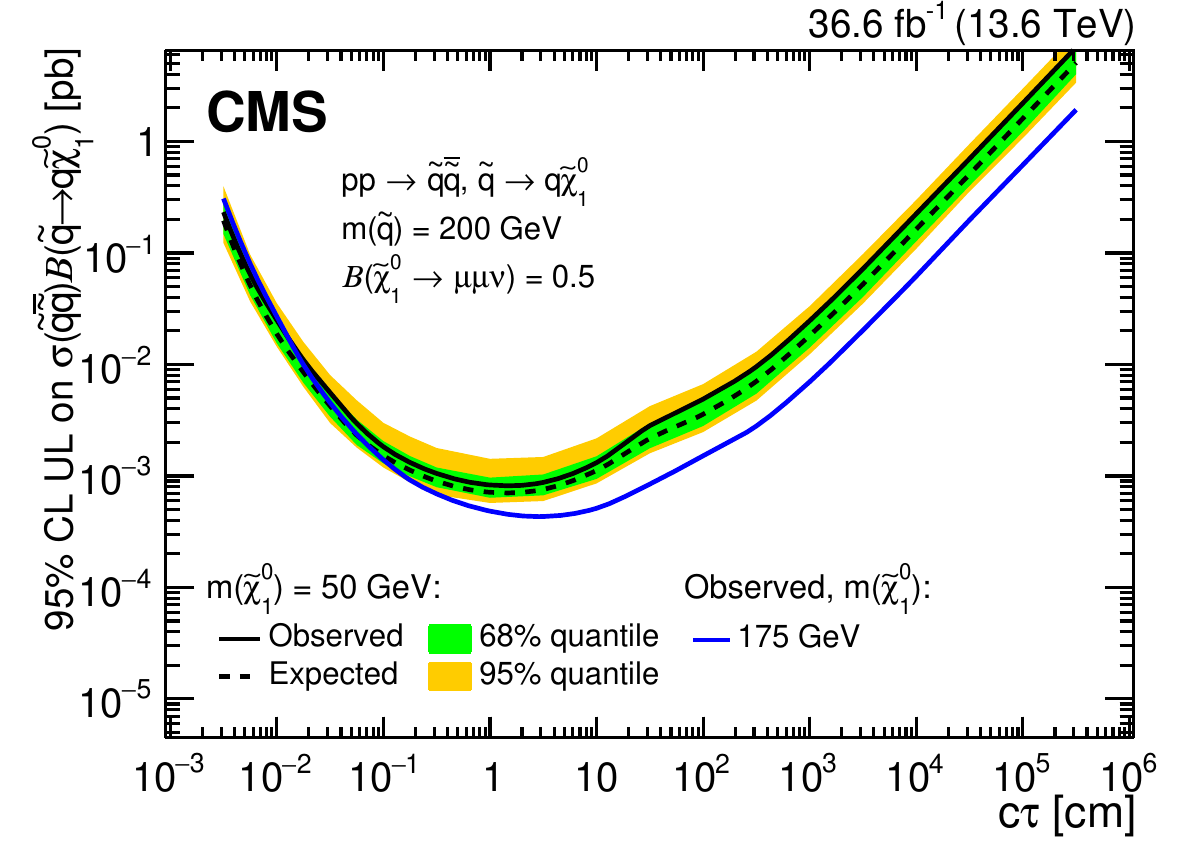}
  \includegraphics[width=0.95\DSquareWidth]{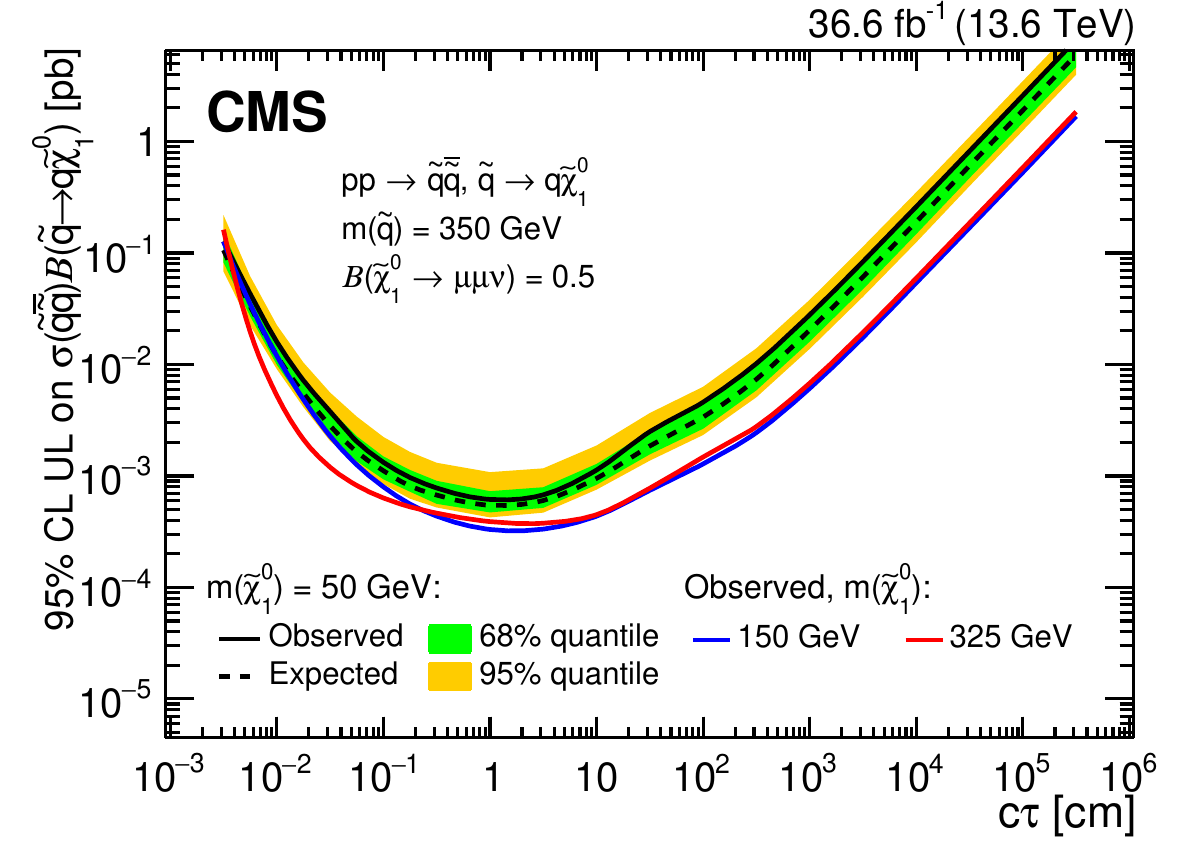}
  \includegraphics[width=0.95\DSquareWidth]{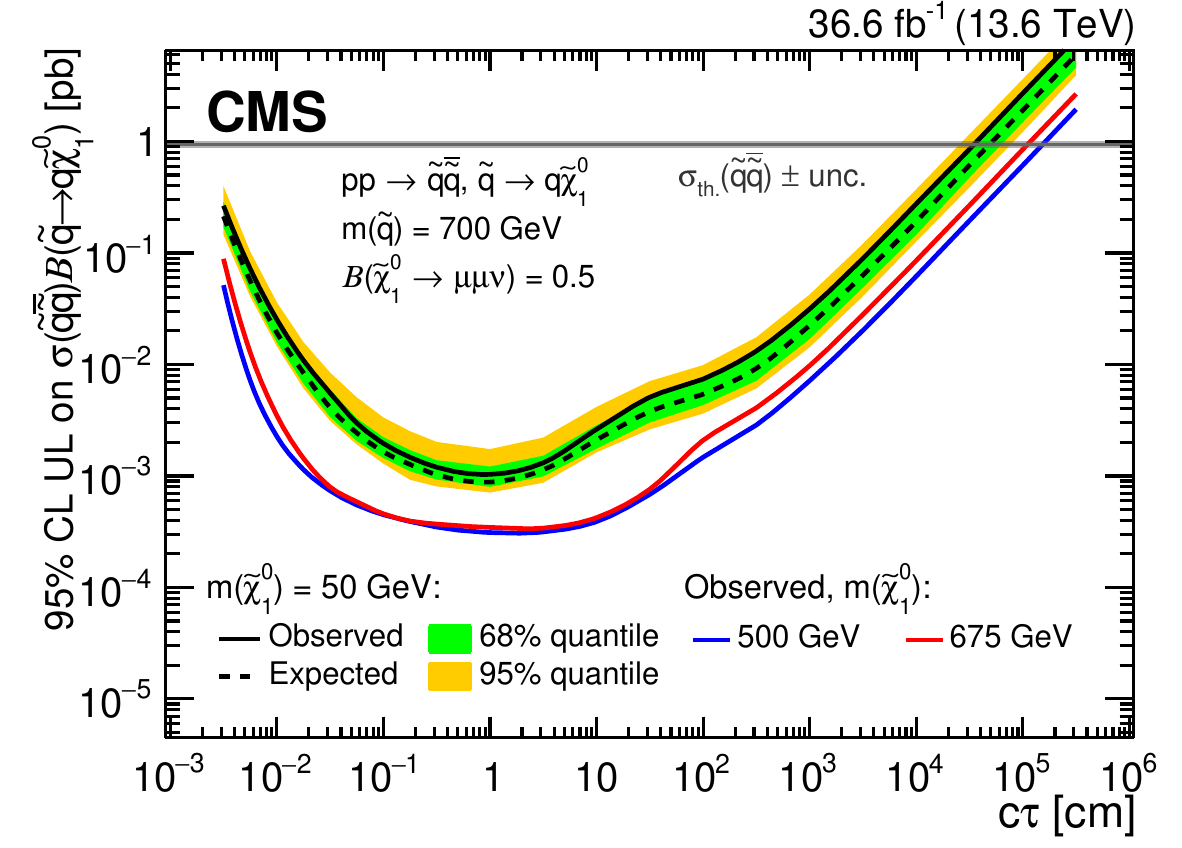}
  \includegraphics[width=0.95\DSquareWidth]{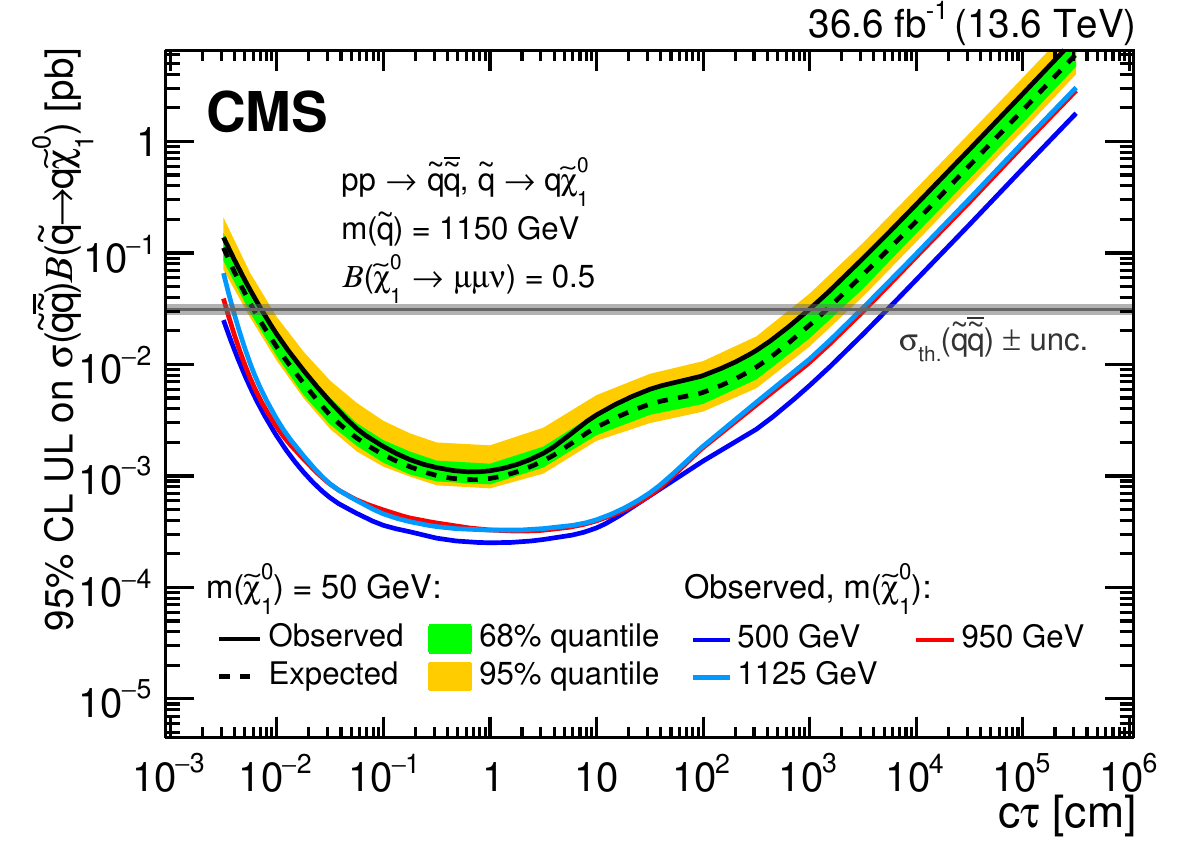}
  \includegraphics[width=0.95\DSquareWidth]{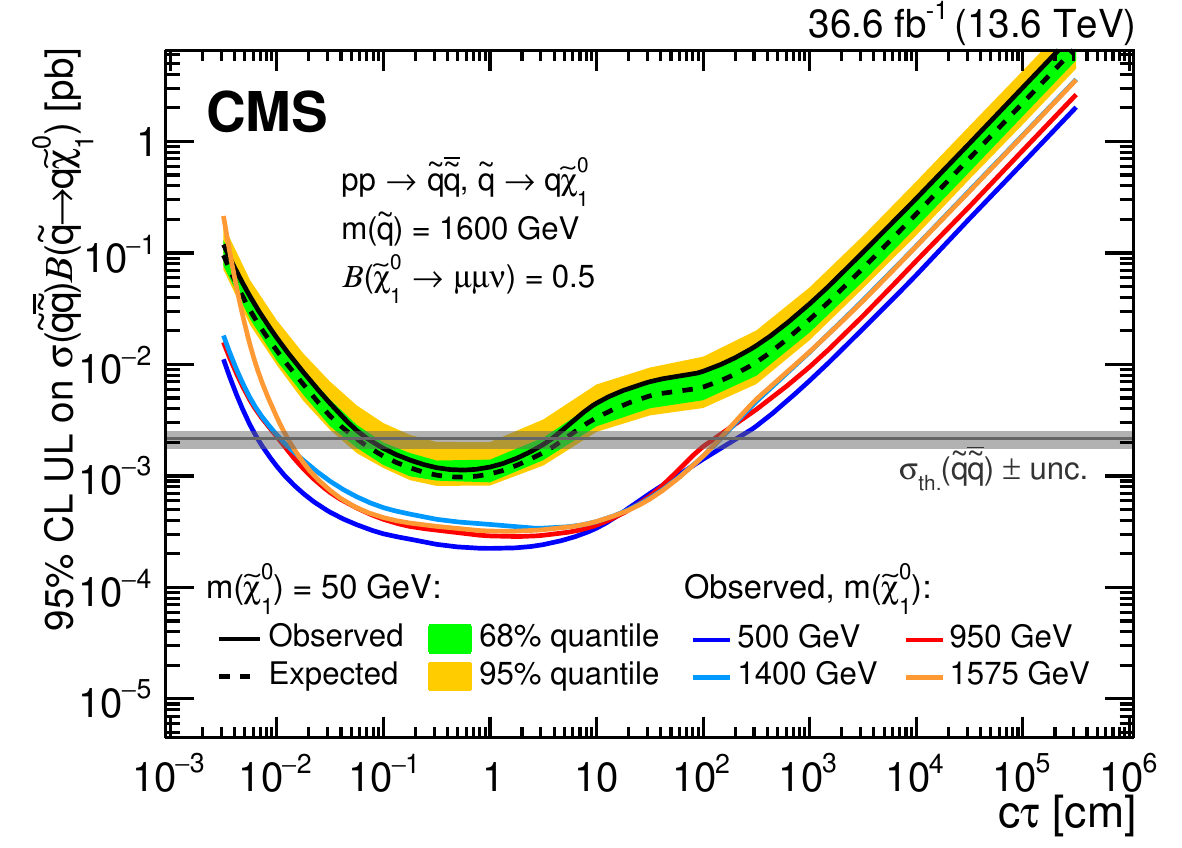}
  \caption{The 95\% \CL upper limits on $\sigma(\Pp\Pp\to\PRPVSquark\PASQ)\mathcal{B}(\squarktoqchi)$ as a
    function of $\cTau(\PRPVChi)$ in the RPV SUSY model, for
    $\mathcal{B}(\chitomumunu) = 0.5$ and $\mSquark$ ranging from
    (upper left) 125\GeV to (lower right) 1.6\TeV.
    The observed limits for various combinations of $\mSquark$ and
    $\mChi$ indicated in the legends are shown as solid curves.  The
    median expected limits and their 68 and 95\% quantiles are shown,
    respectively, as dashed black curves and green and yellow bands
    for the case of $\mChi = 50\GeV$ and omitted for other neutralino
    masses for clarity.  The gray horizontal lines indicate the
    theoretical values of the squark-antisquark production cross
    sections with the uncertainties shown as gray shaded bands.  The
    predicted cross sections for $\mSquark$ = 125, 200, and 350\GeV
    are, respectively, 7200, 840, and 50\unit{pb}, and fall outside
    the $y$-axis range.}
  \label{fig:rpvsusy_limits}
\end{figure}

Figures~\ref{fig:darkphoton_limits} and \ref{fig:darkphoton_limits_Run23}
show the 95\% \CL upper limits obtained in the framework of the HAHM model
under the assumption that $m(\PDarkHiggs) > \mh/2$.
The limits shown in these figures are set on $\mathcal{B}(\PSMHiggs \to \PZD
\PZD)$ as
functions of the mean proper decay length of $\PZD$ for $\mZD$
in the range 10--60\GeV.
Figure~\ref{fig:darkphoton_limits} shows the results obtained in this
analysis, namely the expected limits in the individual dimuon
categories as well as the expected and observed limits obtained in
the ensemble of both categories.
Figure~\ref{fig:darkphoton_limits_Run23} compares the observed limits 
obtained in this analysis with the corresponding Run 2 limits~\cite{EXO-21-006},
and shows the expected and observed limits obtained in a combination
of the two results.

Even though the size of the data sample used by this analysis is about a
factor of 2.5 smaller than that in the Run 2 analysis, the constraints on the
parameters of the HAHM model are comparable or tighter in a
significant fraction of the parameter space, thanks mainly to
improvements in the trigger algorithms.  The combined limits on
$\mathcal{B}(\PSMHiggs \to \PZD \PZD)$ are
approximately a factor of 2 more stringent than the limits obtained in
the Run 2 analysis.
They exclude 
$\mathcal{B}(\PSMHiggs \to \PZD \PZD)$ of 1\% in the range
of $\cTau(\PZD)$ from a few tens of $\mum$ to 30\unit{m} (700\unit{m})
for \mZD = 10\GeV (60\GeV) at 95\% \CL.
In the $\mZD$ interval 20--60\GeV, $\mathcal{B}(\PSMHiggs \to \PZD
\PZD)$ as low as 0.01\% is excluded at 95\% \CL in the $\cTau(\PZD)$
range of about 0.3\mm to 0.5\unit{m}, which corresponds to a wide range
of $\epsilon$ values from ${\approx}10^{-8}$ to ${\approx}10^{-6}$.
These constraints on rare SM Higgs
boson decays are tighter than those derived from searches for
invisible Higgs boson decays~\cite{CMS:2023sdw, Sirunyan:2018owy} and
from measurements of the SM Higgs boson
couplings~\cite{Sirunyan:2018koj}.  At $\mZD> 20\GeV$, the limits
obtained are the most stringent limits to date for all $\cTau(\PZD)$ values except
those between ${\approx}0.1$ and ${\approx}10\cm$ (depending on $\mZD$), where
a previous CMS search~\cite{CMS:2021sch} using data collected with
a dedicated high-rate data stream provides the best limits. 

Figure~\ref{fig:rpvsusy_limits} shows the observed and expected 95\%
\CL upper limits obtained in the ensemble of both dimuon categories in
the framework of the RPV SUSY model.  The limits are set on the product
$\sigma(\Pp\Pp\to\PRPVSquark\PASQ)\mathcal{B}(\squarktoqchi)$
assuming $\mathcal{B}(\chitomumunu) = 0.5$.
They are shown as functions of the mean proper decay length of
$\PRPVChi$ for the various combinations of $\mSquark$ and $\mChi$
indicated in the legends, for $\mSquark$ values ranging from 125\GeV
to 1.6\TeV, and compared to the theoretical predictions.  At a
squark mass of 700\GeV, the data exclude the mean proper neutralino
decay lengths between $30\mum$ and 350\unit{m} (at $\mChi = 50\GeV$),
and between $30\mum$
and 1.5\unit{km} (at $\mChi = 500\GeV$).  At a squark mass of
1.6\TeV, the excluded $\cTau(\PRPVChi)$ range is between 0.07 and
4\unit{cm} for $\mChi = 50\GeV$ and between $70\mum$ and 2\unit{m} for
$\mChi = 500\GeV$.  The limits obtained in this analysis
are tighter than those derived by CMS in the Run~1
search~\cite{EXO-12-037, CMS-PAS-EXO-14-012} in all of the relevant
($\mSquark$, $\mChi$, $\cTau(\PRPVChi)$) parameter space and are more
stringent at $\cTau(\PRPVChi) \lesssim 1\cm$ and $\cTau(\PRPVChi) \gtrsim
1\unit{m}$ than the limits on the $\lambda_{122}$ RPV coupling set by the
ATLAS Collaboration at $\mSquark$ of 700\GeV and
1.6\TeV~\cite{Aad:2683781}.

\section{Summary} \label{sec:summary}
Data collected by the CMS experiment in proton-proton collisions
at $\sqrt{s} = 13.6\TeV$ in 2022 and corresponding to an
integrated luminosity of 36.6\fbinv have been used to conduct an
inclusive search for long-lived exotic neutral particles
decaying to final states with a pair of oppositely charged muons.  The search
strategy is largely model independent
and is sensitive to a broad range of lifetimes and masses.  No significant
excess of events above the standard model background is observed.  The
results are interpreted as limits on the parameters of the
hidden Abelian Higgs model, in which the Higgs boson $\PSMHiggs$ decays to a pair
of long-lived dark photons $\PZD$, and of an $R$-parity violating supersymmetry model,
in which long-lived neutralinos decay to a pair of muons and a neutrino.

Even though the size of the data sample used by this analysis is about a factor of
2.5 smaller than that used in the previous search for displaced dimuons by the CMS experiment in $\Pp\Pp$
collisions at $\sqrt{s} = 13\TeV$, the constraints on the parameters
of the hidden Abelian Higgs model are comparable or tighter in a significant fraction
of the parameter space, thanks mainly to improvements in the
trigger algorithms.
The combination of the
results of this analysis with the results obtained at $\sqrt{s} =
13\TeV$ improves the constraints on the branching fraction of the Higgs
boson to dark photons, $\mathcal{B}(\PSMHiggs \to \PZD \PZD)$, by
approximately a factor of 2.  In the range 10--60\GeV of the $\PZD$ mass \mZD,
$\mathcal{B}(\PSMHiggs \to \PZD \PZD) =  1\%$
is excluded at 95\% confidence level 
in the range of proper decay length $\cTau(\PZD)$ from a few tens of $\mum$ to
30\unit{m} (700\unit{m}) for \mZD = 10\GeV (60\GeV).  
For 
$\mZD$ greater than 20\GeV and  less than $\mh/2$,
the combined limits provide the most stringent
constraints to date on $\mathcal{B}(\PSMHiggs \to \PZD \PZD)$
for $\cTau(\PZD)$ 
between $30\mum$ and ${\approx}0.1\cm$, and above ${\approx}10\cm$. 
When interpreted in the framework of the $R$-parity violating supersymmetry model
at a squark mass of 1.6\TeV, the results exclude mean proper
neutralino decay lengths between 0.07 and 4\unit{cm} for a 50\GeV
neutralino and between $70\mum$ and 2\unit{m} for a 500\GeV
neutralino.

\begin{acknowledgments}
We congratulate our colleagues in the CERN accelerator departments for the excellent performance of the LHC and thank the technical and administrative staffs at CERN and at other CMS institutes for their contributions to the success of the CMS effort. In addition, we gratefully acknowledge the computing centers and personnel of the Worldwide LHC Computing Grid and other centers for delivering so effectively the computing infrastructure essential to our analyses. Finally, we acknowledge the enduring support for the construction and operation of the LHC, the CMS detector, and the supporting computing infrastructure provided by the following funding agencies: SC (Armenia), BMBWF and FWF (Austria); FNRS and FWO (Belgium); CNPq, CAPES, FAPERJ, FAPERGS, and FAPESP (Brazil); MES and BNSF (Bulgaria); CERN; CAS, MoST, and NSFC (China); MINCIENCIAS (Colombia); MSES and CSF (Croatia); RIF (Cyprus); SENESCYT (Ecuador); ERC PRG, RVTT3 and MoER TK202 (Estonia); Academy of Finland, MEC, and HIP (Finland); CEA and CNRS/IN2P3 (France); SRNSF (Georgia); BMBF, DFG, and HGF (Germany); GSRI (Greece); NKFIH (Hungary); DAE and DST (India); IPM (Iran); SFI (Ireland); INFN (Italy); MSIP and NRF (Republic of Korea); MES (Latvia); LMTLT (Lithuania); MOE and UM (Malaysia); BUAP, CINVESTAV, CONACYT, LNS, SEP, and UASLP-FAI (Mexico); MOS (Montenegro); MBIE (New Zealand); PAEC (Pakistan); MES and NSC (Poland); FCT (Portugal); MESTD (Serbia); MCIN/AEI and PCTI (Spain); MOSTR (Sri Lanka); Swiss Funding Agencies (Switzerland); MST (Taipei); MHESI and NSTDA (Thailand); TUBITAK and TENMAK (Turkey); NASU (Ukraine); STFC (United Kingdom); DOE and NSF (USA).
   
\hyphenation{Rachada-pisek} Individuals have received support from the Marie-Curie program and the European Research Council and Horizon 2020 Grant, contract Nos.\ 675440, 724704, 752730, 758316, 765710, 824093, 101115353, and COST Action CA16108 (European Union); the Leventis Foundation; the Alfred P.\ Sloan Foundation; the Alexander von Humboldt Foundation; the Science Committee, project no. 22rl-037 (Armenia); the Belgian Federal Science Policy Office; the Fonds pour la Formation \`a la Recherche dans l'Industrie et dans l'Agriculture (FRIA-Belgium); the Agentschap voor Innovatie door Wetenschap en Technologie (IWT-Belgium); the F.R.S.-FNRS and FWO (Belgium) under the ``Excellence of Science -- EOS" -- be.h project n.\ 30820817; the Beijing Municipal Science \& Technology Commission, No. Z191100007219010 and Fundamental Research Funds for the Central Universities (China); the Ministry of Education, Youth and Sports (MEYS) of the Czech Republic; the Shota Rustaveli National Science Foundation, grant FR-22-985 (Georgia); the Deutsche Forschungsgemeinschaft (DFG), under Germany's Excellence Strategy -- EXC 2121 ``Quantum Universe" -- 390833306, and under project number 400140256 - GRK2497; the Hellenic Foundation for Research and Innovation (HFRI), Project Number 2288 (Greece); the Hungarian Academy of Sciences, the New National Excellence Program - \'UNKP, the NKFIH research grants K 124845, K 124850, K 128713, K 128786, K 129058, K 131991, K 133046, K 138136, K 143460, K 143477, 2020-2.2.1-ED-2021-00181, and TKP2021-NKTA-64 (Hungary); the Council of Science and Industrial Research, India; ICSC -- National Research Center for High Performance Computing, Big Data and Quantum Computing, funded by the NextGenerationEU program (Italy); the Latvian Council of Science; the Ministry of Education and Science, project no. 2022/WK/14, and the National Science Center, contracts Opus 2021/41/B/ST2/01369 and 2021/43/B/ST2/01552 (Poland); the Funda\c{c}\~ao para a Ci\^encia e a Tecnologia, grant CEECIND/01334/2018 (Portugal); the National Priorities Research Program by Qatar National Research Fund; MCIN/AEI/10.13039/501100011033, ERDF ``a way of making Europe", and the Programa Estatal de Fomento de la Investigaci{\'o}n Cient{\'i}fica y T{\'e}cnica de Excelencia Mar\'{\i}a de Maeztu, grant MDM-2017-0765 and Programa Severo Ochoa del Principado de Asturias (Spain); the Chulalongkorn Academic into Its 2nd Century Project Advancement Project, and the National Science, Research and Innovation Fund via the Program Management Unit for Human Resources \& Institutional Development, Research and Innovation, grant B37G660013 (Thailand); the Kavli Foundation; the Nvidia Corporation; the SuperMicro Corporation; the Welch Foundation, contract C-1845; and the Weston Havens Foundation (USA).
\end{acknowledgments}

\bibliography{auto_generated} 
\cleardoublepage \appendix\section{The CMS Collaboration \label{app:collab}}\begin{sloppypar}\hyphenpenalty=5000\widowpenalty=500\clubpenalty=5000
\cmsinstitute{Yerevan Physics Institute, Yerevan, Armenia}
{\tolerance=6000
A.~Hayrapetyan, A.~Tumasyan\cmsAuthorMark{1}\cmsorcid{0009-0000-0684-6742}
\par}
\cmsinstitute{Institut f\"{u}r Hochenergiephysik, Vienna, Austria}
{\tolerance=6000
W.~Adam\cmsorcid{0000-0001-9099-4341}, J.W.~Andrejkovic, T.~Bergauer\cmsorcid{0000-0002-5786-0293}, S.~Chatterjee\cmsorcid{0000-0003-2660-0349}, K.~Damanakis\cmsorcid{0000-0001-5389-2872}, M.~Dragicevic\cmsorcid{0000-0003-1967-6783}, P.S.~Hussain\cmsorcid{0000-0002-4825-5278}, M.~Jeitler\cmsAuthorMark{2}\cmsorcid{0000-0002-5141-9560}, N.~Krammer\cmsorcid{0000-0002-0548-0985}, A.~Li\cmsorcid{0000-0002-4547-116X}, D.~Liko\cmsorcid{0000-0002-3380-473X}, I.~Mikulec\cmsorcid{0000-0003-0385-2746}, J.~Schieck\cmsAuthorMark{2}\cmsorcid{0000-0002-1058-8093}, R.~Sch\"{o}fbeck\cmsorcid{0000-0002-2332-8784}, D.~Schwarz\cmsorcid{0000-0002-3821-7331}, M.~Sonawane\cmsorcid{0000-0003-0510-7010}, S.~Templ\cmsorcid{0000-0003-3137-5692}, W.~Waltenberger\cmsorcid{0000-0002-6215-7228}, C.-E.~Wulz\cmsAuthorMark{2}\cmsorcid{0000-0001-9226-5812}
% ...
\par}
\vskip\cmsinstskip
\dag:~Deceased\\
$^{1}$Also at Yerevan State University, Yerevan, Armenia\\
$^{2}$Also at TU Wien, Vienna, Austria\\
% ...
\end{sloppypar}
\end{document}